\documentclass[twocolumn, 10pt, journal]{IEEEtran}
\usepackage{amsmath}
\usepackage[pdftex]{graphicx}
\usepackage{graphicx}
\usepackage{pstricks}
\usepackage{amsfonts}
\usepackage{tabularx}
\newtheorem{Theorem}{Theorem}
\newtheorem{Lemma}{Lemma}
\newtheorem{Remark}{Remark}
\usepackage{amssymb}
\usepackage{amsfonts}
\usepackage{multirow}
\usepackage{tabularx}
\usepackage{epsfig}
\usepackage{epstopdf}

\usepackage{algorithm,algpseudocode,graphicx}

\begin{document}
\title{Novel Outage-Aware NOMA Protocol for Secrecy Fairness Maximization Among Untrusted Users}
\author{Sapna Thapar,~\IEEEmembership{Student Member,~IEEE,}  Deepak Mishra,~\IEEEmembership{Member,~IEEE,} 
\\and Ravikant Saini,~\IEEEmembership{Member,~IEEE}

\thanks{Copyright (c) 2015 IEEE. Personal use of this material is permitted. However, permission to use this material for any other purposes must be obtained from the IEEE by sending a request to pubs-permissions@ieee.org.}
\thanks{S. Thapar and R. Saini are with the Department of Electrical Engineering, Indian Institute of Technology Jammu, Jammu, Jammu $\&$ Kashmir 181 221, India (e-mail: 2018ree0019@iitjammu.ac.in; ravikant.saini@iitjammu.ac.in). D. Mishra is with the School of Electrical Engineering and Telecommunications, University of New South Wales Sydney, NSW 2052, Australia (e-mail: d.mishra@unsw.edu.au).}
\thanks{This research work was supported by the Tata Consultancy Services Research Scholar Program (TCS RSP) Fellowship.}
\thanks{A preliminary version \cite{globecom} of this work has been published at IEEE GLOBECOM, Hawaii, USA, Dec. 2019.}}
 \maketitle
\begin{abstract}
Observing the significance of spectrally-efficient secure non-orthogonal multiple access (NOMA), this paper proposes a novel quality of service (QoS) aware secure NOMA protocol that maximizes secrecy fairness among untrusted users. Considering a base station (BS) and two users, a novel decoding order is designed that provides security to both users. With the objective of ensuring secrecy fairness between users, while satisfying their QoS requirements under BS transmit power budget constraint, we explore the problem of minimizing the maximum secrecy outage probability (SOP). Closed-form expression of pair outage probability (POP) and optimal power allocation (PA) minimizing POP are obtained. To analyze secrecy performance, analytical expressions of SOP for both users are derived, and individual SOP minimization problems are solved using concept of generalized-convexity. High signal-to-noise ratio approximation of SOP and asymptotically optimized solution minimizing this approximation is also found. Furthermore, global-optimal solution from secrecy fairness standpoint is obtained at low computational complexity, and tight approximation is derived to get analytical insights. Numerical results present useful insights on globally optimized PA which ensure secrecy fairness and provide performance gain of about $55.12\%$, $69.30\%$, and $19.11\%$, respectively, compared to fixed PA and individual users' optimal PAs. Finally, a tradeoff between secrecy fairness performance and QoS demands is presented.

\textit{Index Terms$-$}
5G communications, non-orthogonal multiple access, physical layer security, power allocation, optimization.
\end{abstract}

\section{Introduction}
Exponentially increasing network traffic, imposed by internet-enabled applications, poses severe challenges of supporting massive connections loaded with high data rate demands for upcoming fifth generation (5G) communication networks. Non-orthogonal multiple access (NOMA) has been recognized as a potential breakthrough because of the possibility of accommodating a number of users within the same subcarrier allocation \cite{ding2017survey}. This exhibits a significant spectral efficiency enhancement compared to traditional orthogonal multiple access techniques. But, the broadcast nature of communication at the transmitter makes NOMA vulnerable to the eavesdropping. Therefore, research on security issues in NOMA networks is of great significance. The amalgamation of NOMA and physical layer security (PLS) has been observed as a promising solution to provide spectrally-efficient secure wireless communication \cite{liu2017enhancing}. Despite merits, the design process has secrecy challenge of wiretapping in the presence of untrusted NOMA users.

\subsection{Related Art}
To address the demand for massive connections, modern wireless networks strive for better spectral efficiency. Motivated by the potential of NOMA in providing spectrally-efficient networks, survey papers \cite{ding2017survey}, \cite{8114722} have summarized recent research contributions in power-domain NOMA. Besides, the broadcast nature of wireless transmission leads to security issues on the information-carrying signal. Triggered by these security concerns, researchers have considered the PLS technique which was first recommended by Wyner \cite{wyner1975wire} as a complement to cryptographic approaches. To highlight the capabilities of PLS, \cite{pls} has provided the latest survey of PLS research on diverse 5G technologies, such as multiple input multiple output, full-duplex, and millimeter wave.   

Even though a great amount of research has been done in NOMA and PLS independently, researchers have recently focused attention on using PLS as a security measure in NOMA. An investigation of PLS in large-scale networks by invoking stochastic geometry has been done in \cite{liu2017enhancing}, where a base station (BS) communicates with randomly distributed NOMA users and eavesdroppers. The authors have proposed a protected zone around BS to maintain an eavesdropper free zone. Secure NOMA transmission for multiple users against eavesdropper in single-input single-output network has been studied in \cite{zhang2016secrecy}, where the authors have derived a closed-form expression of optimal power allocation (PA) that maximizes sum secure rate of NOMA users. In \cite{he2017design}, the authors have investigated the optimal decoding order, individual transmission rates, and PA to each user to design secure NOMA system against external eavesdropper under secrecy outage constraint. The authors have proved that optimal decoding order for secure NOMA scheme against external eavesdropper is same as that for conventional NOMA. Secrecy of a cooperative NOMA system with a decode-and-forward and an amplify-and-forward relay against external eavesdropper has been analyzed in \cite{8245831}. The authors in \cite{8695092} have proposed a NOMA-inspired jamming and relaying scheme for enhancing the PLS of untrusted relay networks. In \cite{8333737}, the authors considered a multiple-input single-output NOMA cognitive relay radio network in the presence of multiple eavesdroppers and proposed an artificial noise (AN) aided cooperative jamming scheme for secure transmission in considered network. 

It is noteworthy that the aforementioned works have focused on securing NOMA against external eavesdropper only. Besides external eavesdroppers, secure data transmission among NOMA users is much more challenging due to the underlying SIC based decoding at receivers. Actually, due to the sharing of same resource blocks among users in NOMA systems, users may not trust each other in real scenario, and hence at least some level of secrecy must be provided against internal eavesdropping. Based on the idea, assuming far users as untrusted, sum secrecy rate of trusted near users is investigated in \cite{7833022} for a multiple-input single-output NOMA system. In \cite{basepaper} also, a NOMA system is considered where near user is assumed to be trusted, whereas far user as untrusted, and SOP of trusted user against untrusted user has been investigated. Note that in \cite{7833022} and \cite{basepaper}, it is assumed that the untrusted far user first decodes its own data and then can access the message of the trusted near user with the aid of the SIC technique.

\subsection{Research Gap and Motivation}
Note that existing works \cite{liu2017enhancing}, \cite{zhang2016secrecy}$-$\cite{8333737} have considered potential PLS techniques such as AN aided strategy \cite{liu2017enhancing}, \cite{8333737}, optimal PA \cite{zhang2016secrecy}, \cite{he2017design}, and cooperative relaying \cite{8245831}, \cite{8695092} to enhance secrecy of NOMA transmission against external eavesdroppers assuming all NOMA users as trusted. Since NOMA users share same resource block and have access to decode the message of other paired user to perform SIC, there exists an inherent security issue in NOMA in the case of untrusted users. Therefore, at least some level of secrecy must be provided against internal eavesdropping. Note that the untrusted scenario is a hostile situation, where users have no mutual trust, and each user focuses on achieving secure communication from BS in the presence of other untrusted users, which leads to more complex and constrained resource allocations \cite{saini2019subcarrier}, \cite{saini2016jammer}. It is noteworthy that the existing works \cite{7833022} and \cite{basepaper} only considered the security problem of trusted near user in a NOMA system against untrusted far untrusted user. However, near user also decodes the data of far user, which causes a serious security risk for far user in the case of untrusted near user. Therefore, from security point of view, it is more desirable to design a secure NOMA system for worst-case scenario, where both near and far users are assumed to be untrusted. Thus, with the objective of designing a smart secure communication network, our work focuses on investigating whether secure communication is possible in NOMA in the presence of untrusted users. 

In this direction, \cite{7997115} considered symbol-level SIC (SLSIC) receiver where other user's signal is demodulated but not decoded to perform SIC. In contrast, codeword-level SIC (CLSIC) receiver exists in literature \cite{7146043}, where other user's signal is demodulated and decoded. Since channel decoding is conducted during signal detection in CLSIC receiver, the probability of successful recovery of other user's signal increases in comparison to SLSIC, and hence, the impact of error propagation can be reduced drastically \cite{7146043}. Nevertheless, securing all users' data in an untrusted scenario from a decoding perspective, which is a better system design aspect, has not been considered yet in literature. 

It also needs to noted that when both users are assumed to be untrusted in NOMA, one might think that if any of these users is capable of decoding other user’s signal then security and privacy of decoded users are compromised. But, decoding of users’ signals can even happen in orthogonal multiple access because of the broadcast nature of wireless channels \cite{8823873}. Actually, we focus on achieving secure communication by utilizing the concept of PLS. According to the definition of PLS, information leakage of the legitimate user may occur only when data rate over the desired link is lesser than that of the potential eavesdropper's link because the secrecy rate for a legitimate user is defined as the difference of the rates when a legitimate user decodes itself, and the rate that another user achieves while decoding data of legitimate user \cite{wyner1975wire}, \cite{pls}. \emph{Therefore, considering the definition of PLS into account for an untrusted NOMA system, our work is focused on designing such a secure NOMA transmission protocol that can be feasible in providing positive secrecy rate for all NOMA users.}

Taking this challenge into account, a decoding order has been proposed for a two-user untrusted NOMA scenario in the preliminary version \cite{globecom} of this work that is feasible in providing positive secrecy rate for both near and far NOMA users. As an extension of  \cite{globecom}, in this work, we study all possible decoding orders for a two user NOMA system including conventional one and investigate the best decoding order under secrecy constraint for both users. Besides, in \cite{globecom}, we have considered ideal SIC receivers in which interference from the decoded user is completely removed while decoding the later user. Though better spectral efficiency can be achieved through perfect SIC, but it is not realistic due to various implementation issues such as decoding error and complexity scaling \cite{8114722}, \cite{8370069}. Therefore, a practical scenario would be, when residual interference from the inaccurately decoded user is considered while decoding later user. \textit{To this end, we study secure NOMA to provide positive secrecy rates at all untrusted users with consideration of imperfect SIC based decoding which, to the best of our knowledge, is an open problem.}

It should be emphasized that, with the vision of ensuring secure communication to all users, secrecy performance of the far user against the untrusted near user, and rather secrecy fairness between these users cannot be ignored. Fulfilling users' quality of service (QoS) requirement is also an essential parameter for spectrally-efficient communication. Also, PA to users may play deciding role in obtaining optimal pair outage and secrecy outage performance of a system. Inspired by these solid observations, this paper focuses on investigating numerical and analytical global-optimal PA solutions for a secure NOMA system with untrusted users from a secrecy fairness maximization point of view while satisfying users' QoS requirements.

\subsection{Key Contributions}
The contributions of this work are summarized as follows:
\begin{itemize}
\item Considering a NOMA system with one BS, two untrusted users and imperfect SIC model, a novel optimal decoding order is proposed that provides security to both users.
\item Pair outage probability (POP), as a QoS measure, has been derived analytically. Optimal PA minimizing POP has been obtained using generalized-convexity of POP.  
\item Analytical expressions of SOPs have been derived for both users. SOP minimization problems have been solved using pseudoconvexity of SOP at both users and optimal PAs are obtained numerically. Asymptotic approximations of SOP, and optimal PA for both users are also derived to get analytical insights.
\item To maximize the secrecy fairness between users while satisfying users' QoS demands, we formulate the problem of minimizing the maximum SOP between users under the POP and PA constraints. Global-optimal PA solution is obtained using a low complexity algorithm. Tight analytical approximation for optimal PA is also derived.
\item Extensive simulations are conducted that validate the accuracy of analysis, present insights on optimal performance, and evaluate performance gains by the proposed solution. Tradeoff between secrecy fairness performance of  system and users' QoS demands is also investigated.
\end{itemize}

\subsection{Organization}
The system model, proposed NOMA protocol and investigation on optimal decoding order have been presented in Section II. In Section III, we provide analysis of pair outage performance of the proposed system and derive an optimal PA policy to minimize POP. In section IV, we present SOP analysis for both users, along with their closed-form asymptotic expressions, and also investigate their optimality. Analysis of optimal PA for secrecy fairness maximization between untrusted users under POP constraint is described in Section V. Performance evaluation of proposed analytical model via simulation results is discussed in Section VI, and Section VII presents concluding remarks and future directions.

\section{Secure NOMA Protocol for Untrusted Users}
In this section, we first present the system model, which is followed by the NOMA transmission protocol. Then, an optimal decoding order to secure untrusted NOMA is investigated. 

\begin{figure}[!t]\label{systemmodel}
\centering
\includegraphics[scale=.45]{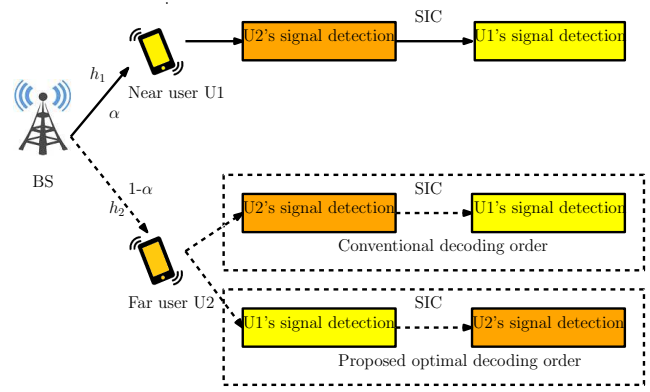}
\caption{Illustration of downlink secure NOMA protocol with proposed optimal decoding order for two untrusted users where decoding order is changed for the far user in comparison to the conventional approach. }
\end{figure}

\subsection{System Model}
We consider the downlink NOMA system with one BS and two untrusted users (Fig. 1). The near and far users are denoted as U$1$ and U$2$, respectively. Our consideration for two-user model is motivated by two reasons: firstly, NOMA system is an interference limited system because co-channel interference is strong in NOMA due to sharing of same resource by multiple users \cite{ding2015cooperative}, \cite{ding2016impact}; and secondly, the implementation complexity at transmitter and receiver sides increases with an increase in number of users because the users perform SIC to remove messages of other users before decoding their own messages \cite{cover2012elements}. Note that due to these reasons, asking all the
users in the system to perform NOMA jointly is not preferable in practice \cite{ding2016impact} and therefore, the users can be divided into multiple groups, where NOMA is implemented within each group. The users are ordered according to their distance to BS. As a result, U$1$ and U$2$ are considered as strong and weak users, respectively \cite{liu2016cooperative}, \cite{do2017bnbf}. The Rayleigh fading channel gain coefficients from BS to U$i$ is denoted by $h_{i}$ for $i \in \{1,2\}$. Channels between BS and users are assumed to be independent and suffer both small scale fading and path loss effects, such that channel power gains $|h_{i}|^{2}$ follows exponential distribution with mean parameter $\lambda_{i}=L_{p}d_{i}^{-n}$. Here $L_{p}$ is path loss constant, $n$ is path loss exponent, and $d_{i}$ is the distance from BS to U$i$. The channel power gains have been sorted such that $|h_{1}|^{2}>|h_{2}|^{2}$ \cite{liu2016cooperative}, \cite{do2017bnbf}. All three nodes are assumed to be equipped with a single antenna \cite{basepaper}. $P_{t}$ denotes total power transmitted by BS to users, and $\alpha$ represents the PA coefficient denoting the fraction of $P_{t}$ allocated to U$1$. Remaining $(1-\alpha)P_t$ is allocated to U$2$.

\subsection{Proposed Secure NOMA Protocol}
In NOMA, the message signals dedicated to the users are superimposed at the BS and then transmitted to the users. The  signal transmitted by BS is $\sqrt{\alpha P_{t}} x_{1} + \sqrt{(1-\alpha)P_{t}} x_{2}$ \cite{basepaper} where $x_{1}$ and $x_{2}$ are unit power signals which contains the message
required by U$1$ and U$2$, respectively. The signals received by U$1$ and U$2$, respectively, from BS are given as 
\begin{equation}
y_{1} = h_{1}( \sqrt{\alpha P_{t}}x_{1} + \sqrt{(1-\alpha)P_{t}} x_{2})  + n_{1},
\end{equation}
\begin{equation}
y_{2} = h_{2}( \sqrt{\alpha P_{t}}x_{1} + \sqrt{(1-\alpha)P_{t}} x_{2})  + n_{2},
\end{equation}
where $n_{1}$ and $n_{2}$ denote zero mean additive white Gaussian noise with variance $\sigma^{2}$ at U$1$ and U$2$, respectively. $\rho_{t}\stackrel{\Delta}{=}P_{t}$/$\sigma^{2}$ denotes BS transmit signal-to-noise ratio (SNR).

After obtaining received signals, two steps are followed at receivers in conventional decoding order: (1) far user decodes its own information signal first and after obtaining its own signal it may decode the information signal of the near user using SIC in case of the far user is an untrusted user \cite{7833022}, \cite{basepaper}; (2) near user first decodes the signal intended for far user to perform SIC, and then decodes its own message. \emph{As a result, near and far users both have access to the signals of far and near user, respectively, which is a critical issue in secure communication with  untrusted users.} Note that the SIC can be performed by all users in a system because SIC on the receivers' side is actually a physical layer capability enabling the receiver to decode packets that arrive collectively. Thus, far user may also decode the information signal of near user using SIC which has been extensively studied in literature also \cite{7833022}, \cite{basepaper}, \cite{7343355}, \cite{9120802}. In this work, we consider imperfect SIC model, where decoding error may occur at both users, resulting in residual interference from the incompletely decoded user after SIC.

Since all users focus to safe their own data from others in untrusted scenario, next we investigate a decoding order that can provide security at both users. Even the conventional decoding order has also not yet been investigated in the literature to ensure positive secrecy for both users in an untrusted NOMA system. Hence, we discuss all possible decoding orders including conventional one to investigate the best decoding order under secrecy constraint. 

\subsection{Optimal Decoding Order from Secrecy Perspective}
In secure NOMA, U$1$ needs to be protected from U$2$ and, vice-versa. Decoding order identifies whether any user decodes its own data first or other user's data. Note that both U$1$ and U$2$ are the multiplexed NOMA users which themselves are the part of the system. Now since BS decides the decoding order for both the users of the system, decoding order can also be changed \cite{7343355}. Thus, for two users' case, four decoding orders are possible. Let us denote the decoding order as $(i,j)$, where $i$ and $j$, respectively, denotes users, i.e., U$i$ and U$j$ (for $i \in \{1,2\}, j \in \{1,2\}$) whose data will be decoded first by U$1$ and U$2$. For example, $(2,1)$ means U$1$ and U$2$ will decode other user's   data first, and then decode its own data. Specifically, the four possible cases are: (a) $(1,1)$, (b) $(1,2)$, (c) $(2,1)$ and (d) $(2,2)$. Under secrecy considerations, the idea is to study whether it is possible to achieve positive secrecy rate at both users. In this regard, investigation of feasible decoding orders out of total possible decoding orders is presented by Theorem $1$.
\begin{Theorem}
\textit{Positive secrecy rate for both near and far users in untrusted NOMA can be obtained in decoding orders $(2,1)$, $(1,2)$, and $(1,1)$.}
\end{Theorem}
\begin{IEEEproof}
We consider each case of decoding orders one by one and investigate the feasibility under positive secrecy constraint. We first consider the conventional decoding order $(2,2)$, where each user first decodes U$2$'s data. Then, the proof continues with a consideration of remaining cases, i.e., $(2,1)$, $(1,2)$ and $(1,1)$. Adopting the imperfect SIC model, we use $\beta_{ij}$ to denote the residual interference from the imperfectly decoded U$i$ when data of U$i$ is decoded by U$j$.

\textit{Case 1: }Considering conventional decoding order $(2,2)$ \cite{basepaper}, U$2$ first decodes its own signal by considering other signal as noise. After decoding its own signal, it may decode the signal of near user using SIC. Then, U$1$ decodes signal associated to U$2$, applies SIC, and then decode its own signal from the remaining signal. Let $\Gamma_{ij}$ denote the received signal-to-interference-plus-noise-ratio at U$i$ when data of U$i$ is decoded by  U$j$. Various $\Gamma_{ij}$ can be given as \cite{basepaper}
\begin{equation}
\Gamma_{21} = \frac{(1-\alpha)| h_{1}|^{2}}{\alpha|h_{1}|^{2} + \frac{1}{\rho_{t}} }, \quad
\Gamma_{11}=\frac{\alpha|h_{1}|^{2}}{\beta_{21}+\frac{1}{\rho_{t}}},
\end{equation}
\begin{equation}
\Gamma_{22} = \frac{(1-\alpha)|h_{2} |^{2}}{\alpha|h_{2}|^{2} + \frac{1}{\rho_{t}} }, \quad
\Gamma_{12} = \frac{\alpha|h_{2}|^{2}}{\beta_{22}+\frac{1}{\rho_{t}}}.
\end{equation}

The achievable secrecy rate $R_{s1}$ of U$1$ can be given as
\begin{equation}\label{Rs1}
R_{s1} = R_{11} - R_{12},
\end{equation}
where $R_{11}$ and $R_{12}$ are given by Shannon's Theorem \cite{tse2005fundamentals} as 
\begin{equation} \label{R11R12}
R_{11} = \log_{2}(1+\Gamma_{11}), \quad
R_{12} = \log_{2}(1+\Gamma_{12}).
\end{equation}

Positive secrecy rate condition $R_{11}> R_{12}$ \eqref{Rs1}, simplified as $\Gamma_{11} > \Gamma_{12}$ \eqref{R11R12} for U$1$ gives a condition $(\rho_{t}\beta_{22}+1)|h_{1}|^{2}>(\rho_{t}\beta_{21}+1)| h_{2}|^{2}$, which shows that based on the values of $\rho_{t}$, $\beta_{22}$, $\beta_{21}$, $|h_{1}|^{2}$ and $|h_{2}|^{2}$, a feasible condition can be obtained. This proves that positive secrecy rate can be achievable at U$1$. Similarly, the achievable secrecy rate $R_{s2}$ of U$2$ is
\begin{equation}
R_{s2} = R_{22} - R_{21}, \label{Rs2}
\end{equation}where
\begin{equation}\label{R22R21}
R_{22} =\log_{2}(1+\Gamma_{22}), \quad
 R_{21} =\log_{2}(1+\Gamma_{21}).
\end{equation}

Here the condition $R_{22} > R_{21}$ \eqref{Rs2}, simplified as $\Gamma_{22}>\Gamma_{21}$ \eqref{R22R21} gives $|h_{2}|^{2} >| h_{1}|^{2}$ which is an infeasible condition because channel power gains are sorted as $|h_{1}|^{2} >| h_{2}|^{2}$. Thus, positive secrecy rate cannot be achieved at U$2$. 

Since our goal is to provide positive secrecy rate to both users, the conventional decoding order cannot be considered for untrusted NOMA because it cannot provide positive secrecy rate for the far user against the untrusted near user. Therefore, next we study other possible decoding orders one by one and investigate the optimal decoding order.

\textit{Case 2: }Next, we consider decoding order $(2,1)$ where both users first decode signals associated to other user, and then decode their own signal after performing SIC. In decoding order (2,1), the order of decoding is different only for the far user in comparison to conventional
decoding order. Note that the far user can even first decode the signal of near user, and then decodes its own signal after applying SIC  \cite{7343355}, \cite{9120802}. As a result, we obtain
\begin{equation}
\Gamma_{21} = \frac{(1-\alpha)|h_{1}|^{2}}{\alpha|h_{1}|^{2}+\frac{1}{\rho_{t}}}, \quad
\Gamma_{11} = \frac{\alpha|h_{1}|^{2}}{\beta_{21}+\frac{1}{\rho_{t}}},
\end{equation}
\begin{equation}
\Gamma_{12} = \frac{\alpha|h_{2}|^{2}}{(1-\alpha)|h_{2}|^{2} + \frac{1}{\rho_{t}}}, \quad
 \Gamma_{22} = \frac{(1-\alpha)|h_{2}|^{2}}{\beta_{12}+\frac{1}{\rho_{t}}}.
\end{equation}

Positive secrecy rate condition $\Gamma_{11}>\Gamma_{12}$, at U$1$ gives
\begin{equation}\label{alpha1}
\alpha < 1 + \frac{|h_{1}|^{2}-|h_{2}|^{2}-|h_{2}|^{2}\rho_{t}\beta_{21}}{|h_{1}|^{2}|h_{2}|^{2}\rho_{t}}.
\end{equation}

Similarly, the feasibility condition $\Gamma_{22}>\Gamma_{21}$ for positive secrecy rate at U$2$, gives  
\begin{equation}\label{alpha2}
\alpha >  \frac{|h_{1}|^{2}-|h_{2}|^{2}+|h_{1}|^{2}\rho_{t}\beta_{12}}{|h_{1}|^{2}|h_{2}|^{2}\rho_{t}}.\\
\end{equation}

Observing \eqref{alpha1} and \eqref{alpha2}, we note that decoding order $(2,1)$ can provide positive secrecy rate to both the users in untrusted NOMA, provided $\alpha$ is constrained as $\left(\frac{|h_{1}|^{2}-|h_{2}|^{2}+|h_{1}|^{2}\rho_{t}\beta_{12}}{|h_{1}|^{2}|h_{2}|^{2}\rho_{t}} < \alpha< 1 + \frac{|h_{1}|^{2}-|h_{2}|^{2}-|h_{2}|^{2}\rho_{t}\beta_{21}}{|h_{1}|^{2}|h_{2}|^{2}\rho_{t}}\right)$.

\textit{Case 3: }In decoding order $(1,2)$, U$1$ and U$2$ first detect their own signals, apply SIC, and then decode signal of other user. Similar to the aforementioned analysis, we obtain $\Gamma_{11}=\frac{\alpha|h_{1}|^{2}}{(1-\alpha)|h_{1}|^{2}+\frac{1}{\rho_{t}}}, \Gamma_{21}=\frac{(1-\alpha)|h_{1}|^{2}}{\beta_{11}+\frac{1}{\rho_{t}}}, \Gamma_{22} = \frac{(1-\alpha)|h_{2}|^{2}}{\alpha|h_{2}|^{2}+\frac{1}{\rho_{t}}},  \Gamma_{12} = \frac{\alpha|h_{2}|^{2}}{\beta_{22}+\frac{1}{\rho_{t}}}$. In this case, $\Gamma_{11}>\Gamma_{12}$ leads to $\alpha > 1 - \frac{|h_{1}|^{2}-|h_{2}|^{2}+|h_{1}|^{2}\rho_{t}\beta_{22}}{|h_{1}|^{2}|h_{2}|^{2}\rho_{t}}$, and $\Gamma_{22} > \Gamma_{21}$ provides $\alpha < \frac{|h_{2}|^{2}-|h_{1}|^{2}+|h_{2}|^{2}\rho_{t}\beta_{11}}{|h_{1}|^{2}|h_{2}|^{2}\rho_{t}}$ which are feasible. Thus, $(1,2)$ is concluded as feasible decoding order in providing positive secrecy rate to both users under constraint $\left(1 - \frac{|h_{1}|^{2}-|h_{2}|^{2}+|h_{1}|^{2}\rho_{t}\beta_{22}}{|h_{1}|^{2}|h_{2}|^{2}\rho_{t}} < \alpha < \frac{|h_{2}|^{2}-|h_{1}|^{2}+|h_{2}|^{2}\rho_{t}\beta_{11}}{|h_{1}|^{2}|h_{2}|^{2}\rho_{t}}\right)$.

\textit{Case 4: }In $(1,1)$ decoding order, U$1$ first decodes its own signal, applies SIC and then decodes signal of other user, whereas U$2$ first decodes signal of other user, subtracts it via SIC and then decodes its own signal. As a result, $\Gamma_{11} = \frac{\alpha|h_{1}|^{2}}{(1-\alpha)|h_{1}|^{2} + \frac{1}{\rho_{t}} }, \Gamma_{21} = \frac{(1-\alpha)|h_{1}|^{2}}{\beta_{11}+\frac{1}{\rho_{t}}}, \Gamma_{12} = \frac{\alpha|h_{2}|^{2}}{(1-\alpha)|h_{2}|^{2} + \frac{1}{\rho_{t}} }, \Gamma_{22} = \frac{(1-\alpha)|h_{2}|^{2}}{\beta_{12}+\frac{1}{\rho_{t}}}$. Here $\Gamma_{11}>\Gamma_{12}$ gives $|h_{1}|^{2}>|h_{2}|^{2}$, and $\Gamma_{22}>\Gamma_{21}$ required for positive secrecy rate at U$2$ gives $(\rho_{t}\beta_{11}+1)|h_{2}|^{2}>(\rho_{t}\beta_{12}+1)|h_{1}|^{2}$ which is also feasible. Thus, $(1,1)$ also can also provide security to both the users, when obtained feasibility conditions are satisfied.
\end{IEEEproof}

Thus, it can be concluded from the aforementioned analysis that utilizing the concept of PLS, three decoding orders, i.e., $(2,1)$, $(1,2)$, $(1,1)$, are feasible in providing positive secrecy rate for both near and far users, but with a suitable constraint on PA. Next, the investigation of optimal decoding order among these three feasible decoding orders is presented through Theorem $2$.  
\begin{Theorem}
\textit{For untrusted NOMA scenario, the optimal decoding order among the feasible decoding orders is $(2,1)$ that gives the best secrecy rate in the system.}
\end{Theorem}
\begin{IEEEproof}
See Appendix A.
\end{IEEEproof}

\begin{Remark}
Note that the number of total possible decoding orders is enormous with more users due to which excessive computational complexity occurs in finding an optimal secure decoding order. Therefore, we have studied the system with two users, however, the investigation can be extended for more users easily.
\end{Remark}

\section{Pair Outage Performance of untrusted NOMA}
In order to ensure users' QoS demands for reliable communication over all the links, we first derive expression of POP for the optimal decoding order $(2,1)$. Next, we obtain optimal PA to minimize POP by utilizing the concept of generalized-convexity. 

\subsection{Pair Outage Probability Analysis}
Each user in a network has a predefined QoS demand, i.e., each user desires the transmitter to send data with a minimum information rate guarantees. POP, denoted as $p_{o}$, ensures minimum rate guarantee to each user. In other words, $p_{o}$ is defined as the probability that a pair outage happens, i.e., the achievable rate at each user falls below a pre-determined threshold rate. Assuming $R_{i}^{th}$ as the threshold rate for U$i$ and defining $\pi_{i}\stackrel{\Delta}{=}2^{R_{i}^{th}}-1$, $p_{o}$ can be given as 
\begin{align}\label{pair_outage}
 p_{o} &= 1 - \text{Pr}\{\Gamma_{11}>\pi_{1}, \Gamma_{21}>\pi_{2}, \Gamma_{12}>\pi_{1}, \Gamma_{22}>\pi_{2}\},\nonumber \\
& \stackrel{(\mathrm{g})}{=} 1 - \text{Pr}\{\Gamma_{11}>\pi_{1}, \Gamma_{21}>\pi_{2}\} \text{Pr}\{ \Gamma_{12}>\pi_{1}, \Gamma_{22}>\pi_{2}\},\nonumber \\ 
& = 1 - \text{Pr}\{|h_{1}|^{2}> \max(\zeta_{1},\zeta_{2})\} \text{Pr}\{|h_{2}|^{2}> \max(\zeta_{3},\zeta_{4})\},\nonumber \\
& = 1 - (1 - F_{|h_{1}|^{2}}(\max(\zeta_{1},\zeta_{2}))) (1 - F_{|h_{2}|^{2}}(\max(\zeta_{3},\zeta_{4}))),\nonumber \\
& = 1 - \bar{F}_{|h_{1}|^{2}}(\max(\zeta_{1},\zeta_{2})) \bar{F}_{|h_{2}|^{2}}(\max(\zeta_{3},\zeta_{4})),
\end{align}
where $\text{Pr}\{.\}$ denotes the probability measure. $(\mathrm{g})$ follows from the property of independent events \cite{papoulis2002probability}. Here $\zeta_{1}\stackrel{\Delta}{=} \frac{\pi_{1}\gamma_{21}}{\rho_{t}\alpha}, \zeta_{2}\stackrel{\Delta}{=}\frac{\pi_{2}}{\rho_{t}(1-\alpha-\alpha\pi_{2})}, \zeta_{3} \stackrel{\Delta}{=}\frac{\pi_{1}}{\rho_{t}(\alpha-(1-\alpha)\pi_{1})}$, and $\zeta_{4} \stackrel{\Delta}{=}\frac{\pi_{2}\gamma_{12}}{\rho_{t}(1-\alpha)}$. Note that $\gamma_{21}\stackrel{\Delta}{=}\rho_{t}\beta_{21}+1$ and $\gamma_{12}\stackrel{\Delta}{=}\rho_{t}\beta_{12}+1$. $F_{|h_{1}|^{2}}(x)$ and $\bar{F}_{|h_{1}|^{2}}(x)$ are cumulative distribution function (CDF) and complementary cumulative distribution function (CCDF), respectively, of channel power gain $|h_{1}|^{2}$. Similarly, $F_{|h_{2}|^{2}}(x)$ and $\bar{F}_{|h_{2}|^{2}}(x)$ are CDF and CCDF, respectively, of channel power gain $|h_{2}|^{2}$. 

Note that $\bar{F}_{|h_{1}|^{2}}(\max(\zeta_{1},\zeta_{2}))$ in \eqref{pair_outage} can be rewritten for two cases $\zeta_{1} >\zeta_{2}$ and $\zeta_{1}<\zeta_{2}$. The first case $\zeta_{1} >\zeta_{2}$, results in a constraint on $\alpha$ as $\alpha < \frac{\pi_{1}\gamma_{21}}{\pi_{1}\gamma_{21}+\pi_{2}+\pi_{1}\pi_{2}\gamma_{21}}$. Similarly, the constraint $\alpha >\frac{\pi_{1}\gamma_{21}}{\pi_{1}\gamma_{21}+\pi_{2}+\pi_{1}\pi_{2}\gamma_{21}}$ results in the second case $\zeta_{1} < \zeta_{2}$. In addition, from the definition of CDF of exponential distribution, we observe that $\zeta_{1}>0$ and $\zeta_{2}>0$ which leads to $\alpha>0$ and $\alpha<\frac{1}{1+\pi_{2}}$, respectively. Considering $\alpha_{1}\!\stackrel{\Delta}{=}\!\frac{\pi_{1}\gamma_{21}}{\pi_{1}\gamma_{21}+\pi_{2}+\pi_{1}\pi_{2}\gamma_{21}}$ and $\alpha_{2}\! \stackrel{\Delta}{=}\!\frac{1}{1+\pi_{2}}$, $\bar{F}_{|h_{1}|^{2}}(\max(\zeta_{1},\zeta_{2}))$ is given as 
\begin{align}\label{pair_outage1}\textstyle
\bar{F}_{|h_{1}|^{2}}(\max(\zeta_{1},\zeta_{2}))\!=\!     
\begin{cases}
 \exp\{-\frac{\zeta_{1}}{\lambda_{1}}\}, \! & 0\!<\!\alpha\!<\!\alpha_{1} \\
 \exp\{-\frac{\zeta_{2}}{\lambda_{1}}\}, \! & \alpha_{1}\!<\!\alpha\!<\!\alpha_{2} \\
  0, & \text{otherwise}.    
\end{cases}  
\end{align}

Similarly, $\bar{F}_{|h_{2}|^{2}}(\max(\zeta_{3},\zeta_{4}))$ can be obtained as follows
\begin{align}\label{pair_outage2}\textstyle
\bar{F}_{|h_{2}|^{2}}(\max(\zeta_{3},\zeta_{4})) \!&=\! 
\begin{cases}
  \exp\{-\frac{\zeta_{3}}{\lambda_{2}}\},  & \alpha_{3}\!<\!\alpha\!<\!\alpha_{4}\\
   \exp\{-\frac{\zeta_{4}}{\lambda_{2}}\}, & \alpha_{4}\!<\!\alpha\!<\!1\\
           0, & \text{otherwise}
      \end{cases}
\end{align}

where $\alpha_{3}=\frac{\pi_{1}}{1+\pi_{1}}$, and $\alpha_{4}=\frac{\pi_{1}+\pi_{1}\pi_{2}\gamma_{12}}{\pi_{1}+\pi_{2}\gamma_{12}+\pi_{1}\pi_{2}\gamma_{12}}$. Using \eqref{pair_outage}, \eqref{pair_outage1}, \eqref{pair_outage2}, the piecewise definition of $p_{o}$ as a function of $\alpha$ is given at the top of next page in \eqref{pair_outage3}
\begin{figure*}
\begin{equation}\label{pair_outage3}
p_{o} = %
\begin{cases}
1-\bar{F}_{|h_{1}|^{2}}(\zeta_{1})  \bar{F}_{|h_{2}|^{2}}(\zeta_{3}), \quad \quad \alpha_{3} < \alpha < \alpha_{1} \\
1-\bar{F}_{|h_{1}|^{2}}(\zeta_{2})  \bar{F}_{|h_{2}|^{2}}(\zeta_{3}), \quad \quad
\begin{cases} 
[\alpha_{1} < \alpha < \alpha_{2}]  \wedge [ \alpha_{2} < \alpha_{4}] \wedge [\alpha_{3} < \alpha_{1}]\\
[\alpha_{1} < \alpha < \alpha_{4}] \wedge [ \alpha_{2} > \alpha_{4}] \wedge [\alpha_{3} < \alpha_{1}]\\
[\alpha_{3} < \alpha < \alpha_{2}] \wedge [ \alpha_{2} < \alpha_{4}] \wedge [\alpha_{3} > \alpha_{1}]\\
[\alpha_{3} < \alpha < \alpha_{4}] \wedge [ \alpha_{2} > \alpha_{4}] \wedge [\alpha_{3} > \alpha_{1}]\\
\end{cases}
\\
1 - \bar{F}_{|h_{1}|^{2}}(\zeta_{2}) \bar{F}_{|h_{2}|^{2}}(\zeta_{4}), \quad \quad \alpha_{4} < \alpha < \alpha_{2}\\
1, \quad \quad \quad \quad \quad \quad \quad \quad \quad \quad \quad \quad \text{otherwise.}
\end{cases}
\end{equation}
\noindent\rule{18cm}{0.5pt}
\end{figure*}
where
\begin{equation}
\bar{F}_{|h_{1}|^{2}}(\zeta_{1}) = \exp\bigg\{-\frac{\pi_{1}\gamma_{21}}{\rho_{t}\alpha\lambda_{1}}\bigg\}, 
\end{equation}
\begin{equation}
\bar{F}_{|h_{1}|^{2}}(\zeta_{2}) = \exp\bigg\{-\frac{\pi_{2}}{\rho_{t}(1-\alpha-\alpha\pi_{2})\lambda_{1}}\bigg\},
\end{equation}
\begin{equation}
\bar{F}_{|h_{2}|^{2}}(\zeta_{3}) = \exp\bigg\{-\frac{\pi_{1}}{\rho_{t}(\alpha-(1-\alpha)\pi_{1})\lambda_{2}}\bigg\}, 
\end{equation}
\begin{equation}
 \bar{F}_{|h_{2}|^{2}}(\zeta_{4}) = \exp\bigg\{-\frac{\pi_{2}\gamma_{12}}{\rho_{t} (1-\alpha)\lambda_{2}}\bigg\}.
\end{equation}

\begin{Remark}From the closed-form expression of $p_{o}$ in \eqref{pair_outage3}, it can be observed that $p_{o}$ depends on key system parameters such as threshold rates of users and SNR. It is noted that POP increases with an increase in threshold rates of users and decreases with an increase in the SNR.
\end{Remark}

\subsection{Pair Outage Probability Minimization}
Observing $p_{o}$ as a function of $\alpha$, the pair outage optimization problem can be formulated as
\begin{align}
(J1): \underset{\alpha}{\text{minimize}} && p_{o}, &&
\text{s.t.} && (C1): 0<\alpha<1,
\end{align}
where $(C1)$ indicates PA coefficient bounds. 

Note that the closed-form piecewise expression of $p_{o}$ in \eqref{pair_outage3} is a function of $\alpha$. Therefore, in order to obtain optimal PA solution, next we need to study each case of piecewise expression one by one and find candidate optimal points to obtain the global-optimal solution. The global-optimal solution is given by Lemma $1$. 
\begin{Lemma}
\textit{The unique global-optimal PA solution $\alpha_{p_{o}}^{*}$ is the feasible optimal point from the set of obtained optimal points that minimizes $p_{o}$, which is given as   
\begin{align}
\!\alpha_{p_{o}}^{*}\!\stackrel{\Delta}{=}\!\underset{\alpha \in \{\alpha_{c1}, \alpha_{r1}, \alpha_{r2}, \alpha_{c2}\}}{\mathrm{argmin}}\!p_{o},
\end{align}}
where $\alpha_{c1}$, $\alpha_{r1}$, $\alpha_{r2}$, and $\alpha_{c2}$ are the candidate optimal points.
\end{Lemma}
\begin{IEEEproof}
In order to solve the optimization problem, closed-form piecewise expression of $p_{o}$ in \eqref{pair_outage3} is considered. Each case is taken one by one to find the global-optimal solution.

\textit{Case 1 :} In the first case, where $p_{o}=1 - \bar{F}_{|h_{1}|^{2}}(\zeta_{1}) \bar{F}_{|h_{2}|^{2}}(\zeta_{3})$ and $\alpha_{3}<\alpha<\alpha_{1}$, first-order derivative obtained by differentiating $p_{o}$ with respect to $\alpha$ is given as
\begin{align}
\frac{\mathrm{d}p_{o}}{\mathrm{d}\alpha}\!=\!-\bigg(\frac{\pi_{1}\gamma_{21}}{\rho_{t}\alpha^2\lambda_{1}} + \frac{\pi_{1}s_{1}}{\rho_{t}(\alpha s_{1}-\pi_{1})^2\lambda_{2}}\bigg)  \nonumber \\
 \exp{\Big\{\frac{-\pi_{1}\gamma_{21}}{\rho_{t}\alpha\lambda_{1}} + \frac{-\pi_{1}}{\rho_{t}(\alpha s_{1}-\pi_{1})\lambda_{2}}\Big \}},
\end{align} 
where $s_{1} \stackrel{\Delta}{=} (1+\pi_{1})$. Since $\frac{\mathrm{d} p_{o}}{\mathrm{d}\alpha}<0$, $p_{o}$ is a monotonically decreasing function of $\alpha$ in this range. The optimal point is taken as the corner, given as $\alpha_{c1} = \alpha_{1}$.
 
\textit{Case 2 :} For second case, where $p_{o}=1 - \bar{F}_{|h_{1}|^{2}}(\zeta_{2}) \bar{F}_{|h_{2}|^{2}}(\zeta_{3})$, the derivative is given as
\begin{align}\label{derivative}
 \frac{\mathrm{d}p_{o}}{\mathrm{d}\alpha}= \bigg(\frac{\pi_{2}s_{2}}{\rho_{t}(1-\alpha s_{2})^2\lambda_{1}} - \frac{\pi_{1} s_{1}}{\rho_{t}(\alpha s_{1}-\pi_{1})^2\lambda_{2}}\bigg)   \nonumber \\
   \exp{\Big\{\frac{-\pi_{2}}{\rho_{t}(1-\alpha s_{2})\lambda_{1}} + \frac{-\pi_{1}}{\rho_{t}(\alpha s_{1}-\pi_{1})\lambda_{2}}\Big\}},
\end{align} 
where $s_{2} \stackrel{\Delta}{=} (1+\pi_{2})$. Note that \eqref{derivative} does not indicate any monotonic behavior. However, we observe that $p_{o}$ is monotonically increasing function of $\alpha$ if $\frac{\pi_{2}s_{2}}{\rho_{t}(1-\alpha s_{2})^2\lambda_{1}}>\frac{\pi_{1}s_{1}}{\rho_{t}(\alpha s_{1}-\pi_{1})^2\lambda_{2}}$, and monotonically decreasing function otherwise. The point of inflection can be obtained by solving $\frac{\pi_{2}s_{2}}{\rho_{t}(1-\alpha s_{2})^2\lambda_{1}} = \frac{\pi_{1}s_{1}}{\rho_{t}(\alpha s_{1}-\pi_{1})^2\lambda_{2}}$ which can be simplified as a quadratic equation $t_{1}\alpha^2 + t_{2}\alpha + t_{3} = 0$ where $t_{1} \stackrel{\Delta}{=} r_{2}s_{1}s_{2} - r_{1}s_{2}, t_{2} \stackrel{\Delta}{=} 2r_{1} - 2r_{2}\pi_{1}, t_{3} \stackrel{\Delta}{=}\pi_{1}^2\pi_{2}s_{2}\lambda_{2} - \pi_{1}s_{1}\lambda_{1}, r_{1} \stackrel{\Delta}{=} \pi_{1}\lambda_{1}s_{1}s_{2},$ and $r_{2} \stackrel{\Delta}{=} \pi_{2}\lambda_{2}s_{1}s_{2}$. Optimal solutions are the roots of the quadratic equation, given as
\begin{equation}\textstyle
 \alpha_{r1},\alpha_{r2}\!=\!\frac{(r_{2}\pi_{1} - r_{1}) \pm \sqrt{(s_{2}\pi_{1}-s_{1})^2\lambda_{1}\lambda_{2}s_{1}s_{2}\pi_{1}\pi_{2}}}{\lambda_{2}s_{1}^2s_{2}\pi_{2}-\lambda_{1}s_{1}s_{2}^2\pi_{1}}.
\end{equation}

\textit{Case 3 :} In the third case, where $p_{o}=1 - \bar{F}_{|h_{1}|^{2}}(\zeta_{2}) \bar{F}_{|h_{2}|^{2}}(\zeta_{4})$, the derivative is given as
\begin{align}
 \frac{\mathrm{d}p_{o}}{\mathrm{d}\alpha} = \bigg(\frac{\pi_{2}s_{2}}{\rho_{t}(1-\alpha s_{2})^2\lambda_{1}} + \frac{\pi_{2}\gamma_{12}}{\rho_{t}(1-\alpha)^2\lambda_{2}}\bigg)  \nonumber \\
  \exp{\Big\{\frac{-\pi_{2}}{\rho_{t}(1-\alpha s_{2})\lambda_{1}} + \frac{-\pi_{2}\gamma_{12}}{\rho_{t}(1-\alpha)\lambda_{2}}\Big\}},
\end{align} 
which is always greater than zero. Thus, $p_{o}$ is a monotonically increasing function of $\alpha$. The optimal point, in this case, is given by the corner point of considered range as $\alpha_{c2} = \alpha_{4}$.

\textit{Case 4 :} In fourth case, $p_{o}$ is a constant, i.e., $p_{o}=1$, and it is the maximum feasible value. 

From above analysis, we observe that $\alpha_{c1}$ and $\alpha_{c2}$ are two corner points due to monotonically decreasing and increasing property, respectively. $\alpha_{r1}$ and $\alpha_{r2}$ are roots as explained in Case $2$. As a result, POP minimization problem has global-optimal solution $\alpha_{p_{o}}^{*}$, which is feasible optimal point from set \{$ \alpha_{c1}$, $\alpha_{r1}$, $\alpha_{r2}$, $\alpha_{c2}$\} at which $p_{o}$ is minimum. 
\end{IEEEproof}

\section{Secrecy Outage Performance}
To analyze secrecy performance, we derive expressions of SOP for both users and investigate optimal PAs for optimizing SOPs. Next, we provide closed-form asymptotic expressions of SOPs, and optimal PAs to gain analytical insights.

\subsection{Secrecy Outage Probability Analysis}
For each user, an outage event happens when the received secrecy rate is below a pre-determined threshold. Particularly, SOP is defined as the probability that the maximum achievable secrecy rate is less than a target secrecy rate. Next, we derive analytical expressions of SOPs for both near and far users which have been denoted by $s_{oi}$ for U$i$.
\subsubsection{Near user} With $R_{s1}$ and $R_{s1}^{th}$ as achievable and target secrecy rate for U$1$, $s_{o1}$ is stated as
\begin{align}\label{SOP_1}
s_{o1} &= \text{Pr}\{R_{s1} <  R_{s1}^{th}\}\nonumber 
=\text{Pr}\Big\{\frac{1+\Gamma_{11}}{1+\Gamma_{12}} < \Pi_{1}\Big\},\nonumber \\
&=\text{Pr}\Big\{|h_{1}|^{2} < \frac{\gamma_{21}\Pi_{1}|h_{2}|^{2}}{\rho_{t}(1-\alpha)|h_{2}|^{2} + 1} + A_{1}\Big\},\nonumber \\
&= \int_{0}^{\infty} F_{\mid h_{1} \mid ^{2}}\bigg(\frac{\gamma_{21}\Pi_{1}| h_{2}|^{2}}{\rho_{t}(1-\alpha) |h_{2}|^{2}+1}+A_{1}\bigg)f_{| h_{2} |^{2}}(y_{1}) dy_{1},\nonumber \\
&=\!1-\! \frac{1}{\lambda_{2}} \int_{0}^{\infty} \!\exp\bigg\{\!\frac{-\gamma_{21}\Pi_{1}y_{1}}{(\rho_{t}(1-\alpha) y_{1}+1)\lambda_{1}}\! - \!\frac{y_{1}}{\lambda_{2}} - \frac{A_{1}}{\lambda_{1}} \bigg\}dy_{1},
\end{align}
where $\Pi_{1} \stackrel{\Delta}{=} 2^{R_{s1}^{th}}$, $A_{1} \stackrel{\Delta}{=}\frac{\gamma_{21}(\Pi_{1}-1)}{\rho_{t}\alpha}$, and $f_{\mid h_{2} \mid ^{2}}(x)$ is probability density function (PDF) of channel power gain $|h_{2}|^{2}$.
\subsubsection{Far user} With $R_{s2}$ and $R_{s2}^{th}$ as achievable and target secrecy rate for U$2$, $s_{o2}$ is given as
\begin{align}\label{SOP_2}
s_{o2}  &= \text{Pr}\{R_{s2} <  R_{s2}^{th}\} \nonumber 
=\text{Pr}\Big\{\frac{1+\Gamma_{22}}{1+\Gamma_{21}} < \Pi_{2}\Big\}, \nonumber \\
&= \text{Pr}\Big\{|h_{2}|^{2} < \frac{\gamma_{12}\Pi_{2}| h_{1} | ^{2}}{\rho_{t}\alpha| h_{1} |^{2} + 1} + A_{2}\Big\}, \nonumber \\
&= \int_{0}^{\infty} F_{\mid h_{2} \mid ^{2}}\bigg(\frac{\gamma_{12}\Pi_{2}|h_{1}|^{2}}{\rho_{t}\alpha|h_{1} |^{2}+1}+A_{2}\bigg)f_{| h_{1} |^{2}}(y_{2}) dy_{2}, \nonumber \\
&=\!1\!-\!\frac{1}{\lambda_{1}}\int_{0}^{\infty} \exp\bigg\{\frac{-\gamma_{12}\Pi_{2}y_{2}}{(\rho_{t}\alpha y_{2}+1)\lambda_{2}} - \frac{y_{2}}{\lambda_{1}}-\frac{A_{2}}{\lambda_{2}} \bigg\} dy_{2},
\end{align}
where $\Pi_{2} \stackrel{\Delta}{=} 2^{R_{s2}^{th}}$, $A_{2} \stackrel{\Delta}{=} \frac{\gamma_{12}(\Pi_{2}-1)}{\rho_{t}(1-\alpha)}$, and $f_{| h_{1} |^{2}}(x)$ denotes PDF of channel power gain $| h_{1}| ^{2}$.

\subsection{Individual Secrecy Outage Probability Minimization}\label{individual_sop}
Next, we investigate optimality of SOPs of both the users. 
\subsubsection{Near User}
Considering $s_{o1}$ derived in \eqref{SOP_1} as a function of $\alpha$, SOP minimization problem for U$1$ can be formulated as 
\begin{equation} \label{problem_form_nearuser}
\begin{aligned}
(J2): \underset{\alpha}{\text{minimize}}
&&  s_{o1}, 
&& \text{s.t.} &&  (C1).\\
\end{aligned}
\end{equation}

Noting the complexity of derived expression of $s_{o1}$, next we solve SOP minimization problems using pseudoconvexity of SOP. The feasibility of unique solution is asserted by Lemma $2$. 
\begin{Lemma}
\textit{$s_{o1}$ is a pseudoconvex function of $\alpha$.} 
\end{Lemma}
\begin{IEEEproof}
$s_{o1}$ in \eqref{SOP_1} can be rewritten as
\begin{align}\label{SOP11}\textstyle
s_{o1}\!=\!1\!-\!\frac{1}{\lambda_{2}}\int\limits_{0}^{\infty}\exp\bigg\{\!-\!\frac{\gamma_{21}\Pi_{1}y_{1}}{(\rho_{t}(1-\alpha)y_{1}+1)\lambda_{1}}\!-\!\frac{y_{1}}{\lambda_{2}}\!-\!\frac{\gamma_{21}(\Pi_{1}-1)}{\rho_{t}\alpha\lambda_{1}}\bigg\}dy_{1}.
\end{align}

Denoting the integrand of $s_{o1}$ as $I_{1}$, it can be defined as
\begin{equation}\textstyle
I_{1} = \exp\bigg\{-\frac{\gamma_{21}\Pi_{1}y_{1}}{(\rho_{t}(1-\alpha) y_{1}+1)\lambda_{1}} - \frac{y_{1}}{\lambda_{2}} -\frac{\gamma_{21}(\Pi_{1}-1)}{\rho_{t}\alpha \lambda_{1}}\bigg\}.
\end{equation}

Second-order derivative of the logarithm of the integrand function $I_{1}$, with respect to $\alpha$ is  
\begin{equation}
\begin{split}\textstyle
 \frac{\mathrm{d}^{2}\log(I_{1})}{\mathrm{d}\alpha^{2}} = - \Bigg(\frac{2\gamma_{21} (\Pi_{1}-1)}{\rho_{t}\lambda_{1}\alpha^{3}} +
 \frac{2\gamma_{21}\Pi_{1}\rho_{t}^{2} y_{1}^{3}}{\lambda_{1} ((1-\alpha)\rho_{t} y_{1} + 1)^{3}}\Bigg),
\end{split}
\end{equation}
which is non-increasing. This indicates that the integrand $I_{1}$ of the objective function $s_{o1}$ is a logarithmically concave (log-concave) function. Since log-concavity is preserved under integration \cite{boyd2004convex}, we note that integral function in \eqref{SOP11} is also log-concave function. Observing pseudoconcave property \cite[Lemma 5]{mishra2016joint} of log-concave function, we can state that integral function of \eqref{SOP11} is pseudoconcave, and negative of a pseudoconcave function is a pseudoconvex function  \cite{bazaara1979nonlinear}. Hence, $s_{o1}$ is a pseudoconvex function of $\alpha$. 
\end{IEEEproof} 

Due to pseudoconvexity of $s_{o1}$ of U$1$, there exists unique PA that minimizes $s_{o1}$ \cite[Chap. 3.5.9]{bazaara1979nonlinear}. We apply a computationally efficient golden section search algorithm \cite{chang2009n} to find $\alpha_{1}^{*}$ (optimal value of $\alpha$). The algorithm considers pseudoconvex function $s_{o1}$, $\alpha_{lb}$ and $\alpha_{ub}$ as input, where $\alpha_{lb}$ and $\alpha_{ub}$, respectively, are the lower and upper bounds of $\alpha$ at initial stage. It gives optimal solution $\alpha_{1}^{*}$ and corresponding minimum SOP as the output. Since $(0<\alpha<1)$, we first consider $\alpha_{lb}=0$ and $\alpha_{ub}=1$ and then the algorithm searches along $\alpha$ with acceptable tolerance level of $\epsilon<<1$. The algorithm works by reducing the search space with a fixed ratio of $0.618$ at the end of every iteration. Finally, the algorithm terminates after $N$ iterations if search length is less than the given tolerance \cite{chang2009n}. 
\subsubsection{Far User}
SOP minimization problem for U$2$, considering expression of $s_{o2}$ \eqref{SOP_2} is stated as 
\begin{equation} \label{problem_form_faruser}
\begin{aligned}
(J3):\underset{\alpha}{\text{minimize}}
&&  s_{o2}, 
&& \text{s.t.} &&  (C1).\\
\end{aligned}
\end{equation}

In Lemma $3$, we prove the feasibility of unique solution. 
\begin{Lemma}
\textit{$s_{o2}$ is a pseudoconvex function of $\alpha$.} 
\end{Lemma}
\begin{IEEEproof}
Similar to the proof of $s_{o1}$ minimization problem, the pseudoconvexity of SOP for U$2$ can be proved. 
\end{IEEEproof}

Here also, optimal PA solution of $s_{o2}$ minimization problem can be evaluated numerically using golden section search algorithm. Now, the algorithm takes pseudoconvex function $s_{o2}$, $\alpha_{lb}=0$, $\alpha_{ub}=1$ as input. It provides $\alpha_{2}^{*}$, and corresponding minimum SOP as optimal output. 

\subsection{Asymptotic SOP: Analysis and Optimization}\label{individual_sop_asymp}
Observing the complexity of derived expressions, and involved computations in optimizing SOPs, optimal PAs have been obtained numerically. In order to gain analytical insights, next we derive tight asymptotic expressions of SOPs and optimal PAs under high SNR scenario. 
\subsubsection{Near User}
Asymptotic approximation of SOP for U$1$, $\hat s_{o1}$, can be obtained by setting $\rho_{t}\gg1$ in \eqref{SOP11}, which leads to $(\rho_{t}(1-\alpha) y_{1}+1)$ $\approx$ $\rho_{t}(1-\alpha)y_{1}$. Thus, $\hat s_{o1}$ is given as
\begin{align} \label{sop1_asy}\textstyle
\hat s_{o1}\!&=\!1\!-\!\exp\bigg\{\!\frac{-\gamma_{21}\Pi_{1}}{\rho_{t}(1\!-\!\alpha)\lambda_{1}}\!-\!\frac{\gamma_{21}(\Pi_{1}\!-\!1)}{\rho_{t}\alpha\lambda_{1}}\!\bigg\}\!\int_{0}^{\infty}\!\frac{\exp\{\frac{-y_{1}}{\lambda_{2}}\}}{\lambda_{2}}dy_{1}, \nonumber \\
\!&=\!1\!-\!\exp\bigg\{\frac{\gamma_{21}(\Pi_{1}+\alpha-1)} {\rho_{t}\alpha(\alpha-1)\lambda_{1}}\bigg\}.
\end{align}

Considering \eqref{sop1_asy}, $\hat s_{o1}$ minimization problem is given as 
\begin{equation}
\begin{aligned}
(J4): \underset{\alpha}{\text{minimize}}
&&  \hat s_{o1}, 
&& \text{s.t.} && (C1).\\
\end{aligned}
\end{equation}

Asymptotic optimal PA for $(J4)$ is given by Lemma $4$. 
\begin{Lemma}
\textit{The optimal PA $\hat \alpha_{1}$ minimizing $\hat s_{o1}$, is given as
\begin{equation}\label{optimal_a_asy_sop1}
\hat \alpha_{1} = - (\Pi_{1}-1) + \sqrt{(\Pi_{1}(\Pi_{1} -1)}.
\end{equation}}
\end{Lemma}
\begin{IEEEproof}
See Appendix B.
\end{IEEEproof}

\subsubsection{Far User}
Similarly, the asymptotic expression $\hat s_{o2}$ for U$2$, using $(\rho_{t}\alpha y_{2}+1)$ $\approx$ $\rho_{t}\alpha y_{2}$ in \eqref{SOP_2}, can be given as
\begin{align}\label{sop2_asy}\textstyle
\hat s_{o2}\!&=\!1\!-\!\exp\bigg\{ \frac{-\gamma_{12} \Pi_{2}}{\rho_{t}\alpha\lambda_{2}} - \frac{\gamma_{12}(\Pi_{2}-1)}{\rho_{t}(1-\alpha)\lambda_{2}} \bigg\} \int_{0}^{\infty}\frac{\exp\{\frac{-y_{2}}{\lambda_{1}}\}}{\lambda_{1}}dy_{2}, \nonumber \\
\!&=\!1\!-\!\exp\bigg\{\frac{\gamma_{12}(\Pi_{2}-\alpha)} {\rho_{t}\alpha(\alpha-1)\lambda_{2}}\bigg\}.
\end{align}

The $\hat s_{o2}$ \eqref{sop2_asy} minimization problem for U$2$ can be stated as 
\begin{equation}
\begin{aligned}
(J5):\underset{\alpha}{\text{minimize}}
&&  \hat s_{o2}, 
&& \text{s.t.} && (C1).\\
\end{aligned}
\end{equation}
\begin{Lemma}
\textit{The unique PA $\hat \alpha_{2}$ minimizing $\hat s_{o2}$ is given as
\begin{equation}\label{optimal_a_asy_sop2}
\hat \alpha_{2} =  \Pi_{2} - \sqrt{(\Pi_{2}(\Pi_{2} -1)}.
\end{equation}}
\end{Lemma}
\begin{IEEEproof}
See Appendix C.
\end{IEEEproof} 

\begin{Remark}
It can be observed from the asymptotic approximations of SOPs, i.e., $\hat s_{o1}$ and $\hat s_{o2}$, respectively, for U$1$ and U$2$, that the SOPs for near user and far user depend on target secrecy rates and SNR. Note that SOP increases with an increase in target secrecy rate of the user and decreases with an increase in the SNR.
\end{Remark}

\section{Secrecy Fairness Maximization}
In order to ensure secrecy rate guarantee to both the users while satisfying users' QoS requirements, next we formulate secrecy fairness optimization problem and investigate optimal PA. Also, we present asymptotic approximation of optimal PA to get deeper insights. 

\subsection{Problem Formulation}
QoS demands of users, in the form of maximum allowable POP forces an upper bound $\xi$ on $p_{o}$, i.e., $p_{o} \leq \xi$. Hence, QoS constrained secrecy fairness maximization problem using \eqref{pair_outage3}, \eqref{SOP_1} and \eqref{SOP_2} that minimizes the maximum SOP between users under power budget and POP constraints can be formulated as
\begin{align}\label{problem_formulation_qos}
(J6): \underset{\alpha}{\text{minimize }}  \max[s_{o1},s_{o2}], \nonumber
\\  \text{s.t.} \quad (C1),  (C2): p_{o} \leq \xi. 
\end{align}

Using $x_{c} \stackrel{\Delta}{=} \max[s_{o1},s_{o2}]$, an equivalent formulation of $(J6)$ can be obtained as follows
\begin{align}\label{problem_formulation1_qos}
(J7): \underset{\alpha, x_{c}}{\text{minimize}} \quad x_{c}, \quad
\text{s.t.} \quad (C1), (C2), \nonumber
\\ (C3): s_{o1}\leq x_{c},  (C4): s_{o2}\leq x_{c},
\end{align}
where $(C3)$ and $(C4)$ appears from the definition of max$[\cdot]$. 

Due to the presence of nonconvex constraints $(C3)$ and $(C4)$, $(J7)$ is a nonconvex problem. We solve it by analyzing candidate optimal points that are characterized by Karush-Kuhn-Tucker (KKT) conditions \cite{ravindran2006engineering}.

\subsection{Power Control for optimizing Min-Max Secrecy Outage}
First, we solve min-max SOP optimization problem without consideration of users' rate demands. Next, QoS demand is taken into account and optimal NOMA protocol is analyzed. 

\subsubsection{Without Pair Outage Constraint}
The secrecy fairness maximization problem without $(C2)$ can be formulated as
\begin{align}\label{problem_formulation_no_qos}
(J8): \underset{\alpha, x_{c}}{\text{minimize }} && x_{c}, && \text{s.t.} && (C1), (C3), (C4). 
\end{align}

The global-optimal solution is given by Lemma $6$.
\begin{Lemma}
\textit{The global-optimal solution $\widetilde{\alpha_{sop}}$ of the nonconvex problem $(J8)$, at which maximum SOP between users is minimized, is given as
\begin{equation}\label{optimal_a_asy_sop}
\!\widetilde{\alpha_{sop}} \! \stackrel{\Delta}{=}\!\underset{\alpha \in \{\alpha_{1}^{*}, \alpha_{2}^{*}, \alpha_{3}^{*}\}}{\mathrm{argmin}}\! \max[s_{o1},s_{o2}],
\end{equation}
where $\alpha_{1}^{*}, \alpha_{2}^{*}, \alpha_{3}^{*}$ are, respectively, obtained by golden section search algorithm by minimizing $s_{o1}, s_{o2}$ (cf. Section \ref{individual_sop}), and solving $s_{o1}=s_{o2}$.}
\end{Lemma}

\begin{IEEEproof}
Keeping the boundary constraint $(C1)$ implicit, and associating Lagrange multipliers $\eta_{1}$ with $(C3)$ and $\eta_{2}$ with $(C4)$, the Lagrangian function $\mathcal{L}$ can be stated as
\begin{equation}\label{lagragian_function1}
\mathcal{L} \stackrel{\Delta}{=}  x_{c} + \eta_{1}[ s_{o1} - x_{c}] + \eta_{2}[ s_{o2} -  x_{c}].
\end{equation}

The corresponding primal feasibility KKT conditions are given by the constraints $(C3)$ and $(C4)$. Following \eqref{problem_formulation_no_qos} and \eqref{lagragian_function1}, the dual feasibility conditions are given as $\eta_{1}\geq 0$ and $\eta_{2}\geq 0$. The subgradient conditions are obtained as
\begin{equation}
\frac{\mathrm{d}\mathcal{L}}{\mathrm{d} x_{c}}  =  1-\eta_{1}-\eta_{2}= 0, 
\quad
\frac{\mathrm{d}\mathcal{L}}{\mathrm{d}\alpha}  =  \eta_{1} \frac{\mathrm{d} s_{o1}}{\mathrm{d}\alpha}  + \eta_{2}\frac{\mathrm{d} s_{o2}}{\mathrm{d}\alpha} = 0.
\end{equation}

The two complementary slackness conditions are given as
\begin{equation}\label{sop1=sop2_1}
 \eta_{1} [ s_{o1} - x_{c}] = 0, \quad
 \eta_{2} [ s_{o2} -  x_{c}] = 0.
\end{equation}

Note that there exists three cases. \emph{Case 1: $\eta_{1} > 0$ and $\eta_{2}= 0$}, implies $\frac{\mathrm{d} s_{o1}}{\mathrm{d}\alpha}=0$ which results in the same solution as SOP minimization problem \eqref{problem_form_nearuser} of U$1$, i.e., $\alpha=\alpha_{1}^{*}$. \emph{Case 2: $\eta_{2} > 0$ and $\eta_{1}= 0$}, implies $\frac{\mathrm{d}s_{o2}}{\mathrm{d}\alpha}=0$, which results in the solution of SOP minimization \eqref{problem_form_faruser} of U$2$, i.e., $\alpha = \alpha_{2}^{*}$. \emph{Case 3: $\eta_{1} > 0$ and $\eta_{2} > 0$}, implies $ s_{o1} = s_{o2}$ using \eqref{sop1=sop2_1}, which indicates equal SOP for both the users, and this case gives $\alpha = \alpha_{3}^{*}$. Hence, $(J8)$ has three optimal points, i.e., $\alpha_{1}^{*}$ and $\alpha_{2}^{*}$ for minimizing $s_{o1}$ and $s_{o2}$, respectively, and $\alpha_{3}^{*}$ is obtained from the condition of $s_{o1}=s_{o2}$. Finally, the global-optimal solution $\widetilde{\alpha_{sop}}$ to this problem is obtained at the optimal point for which maximum SOP between users is minimum.
\end{IEEEproof} 

Thus, we conclude that since the secrecy fairness maximization problem is nonconvex due to the presence of nonconvex constraints, we have successfully obtained all possible optimal points of this problem by KKT conditions which are the candidates for the global-optimal solution \cite{ravindran2006engineering}. Lastly, the optimal point for which maximum SOP between users is minimum is considered as the global-optimal solution.

\subsubsection{With Pair Outage Constraint}
Now we solve $(J7)$ problem by taking QoS requirements into account. In this case, the global-optimal solution from secrecy perspective must satisfy the constraint $(C2)$. Theoretically $\alpha \in [0, 1]$, but due to $(C2)$, acceptable range of $\alpha$ gets limited, as obtained in Lemma $7$. 
\begin{Lemma}
\textit{The lower and upper bound on $\alpha$, required to achieve maximum allowable POP, are given as
\begin{equation}
\alpha_{lb}^{*}=\max[\alpha_{l1}^{*}, \alpha_{l2}^{*}, \alpha_{l3}^{*}], \quad
\alpha_{ub}^{*}=\min[\alpha_{u1}^{*}, \alpha_{u2}^{*}, \alpha_{u3}^{*}],
\end{equation}
respectively, where $\alpha_{l1}^{*}, \alpha_{l2}^{*}, \alpha_{l3}^{*}, \alpha_{u1}^{*}, \alpha_{u2}^{*}, \alpha_{u3}^{*}$ are the roots of quadratic equations obtained by making $(C2)$ an active constraint, i.e., $p_{o}=\xi$}.
\end{Lemma}

\begin{figure}[!t]\label{flowchart}
\centering
\includegraphics[scale=.4]{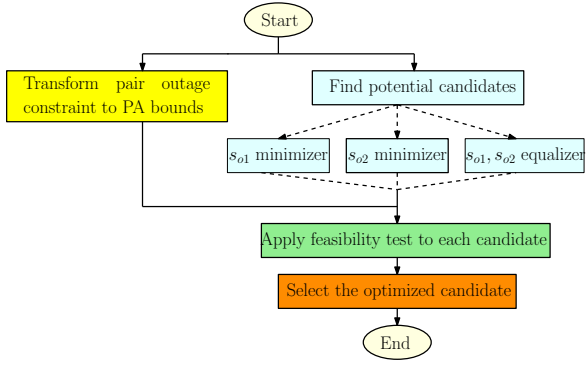}
\caption{Proposed solution methodology for pair outage constrained secrecy fairness maximization.}
\end{figure}

\begin{IEEEproof}
Due to POP constraint, tight analytical bounds on $\alpha$  are obtained by using the expression of $p_{o}$ given in \eqref{pair_outage3}. Since $p_{o}$ is a piecewise function of $\alpha$, we first consider each case of \eqref{pair_outage3} one by one, substitute $p_{o}=\xi$, and find bounds in accordance with the respective $\alpha$ range. In first case of $p_{o}$, where $p_{o}=1 - \bar{F}_{|h_{1}|^{2}}(\zeta_{1}) \times \bar{F}_{|h_{2}|^{2}}(\zeta_{3})$ and $\alpha_{3}< \alpha<\alpha_{1}$, the lower and upper bound can be obtained by solving $p_{o} = \xi$ which results in a quadratic equation given as $\log(1-\xi)s_{1}\alpha^{2}-(\pi_{1}\log(1-\xi)-z_{1}s_{1}-z_{2})\alpha-\pi_{1}z_{1}=0$, where $z_{1}\stackrel{\Delta}{=}\frac{\gamma_{21}\pi_{1}}{\rho_{t}\lambda_{1}}$ and $z_{2}\stackrel{\Delta}{=}\frac{\pi_{1}}{\rho_{t}\lambda_{2}}$. The roots of the quadratic equation, denoted as $\alpha_{l1}$ and $\alpha_{u1}$, are given in \eqref{roots1} at next page. 

\begin{figure*}
\begin{equation}\label{roots1}
\alpha_{u1}, \alpha_{l1}\!=\! \frac{(\pi_{1}\log(1-\xi)-z_{1}s_{1}-z_{2})\!\pm\! \sqrt{(\pi_{1}\log(1-\xi)-z_{1}s_{1}-z_{2})^2\!+\!4 \log(1-\xi)s_{1}\pi_{1}z_{1}}}{2\log(1-\xi)s_{1}}.\\
\end{equation}
\noindent\rule{18cm}{0.5pt}
\end{figure*}

The lower bound of PA in this case can be obtained as $\alpha_{l1}^{*}=\alpha_{l1}$ iff $\alpha_{l1}$ exists as a feasible point in the range else $\alpha_{l1}^{*}=0$. Similarly, the upper bound of PA in this case can be given as $\alpha_{u1}^{*}=\alpha_{u1}$ iff $\alpha_{3}<\alpha_{u1}<\alpha_{1}$ otherwise $\alpha_{u1}^{*}=1$. Similar to the above analysis, lower and upper bounds can be obtained for other cases also. Let us denote lower and upper bounds, respectively, for the second case as $\alpha_{l2}^{*}$ and $\alpha_{u2}^{*}$, and for third case as $\alpha_{l3}^{*}$ and $\alpha_{u3}^{*}$. After identifying lower and upper bounds of PA in each case, the $\alpha_{lb}^{*}$ in the whole range can be obtained as $\alpha_{lb}^{*}=\max[\alpha_{l1}^{*}, \alpha_{l2}^{*}, \alpha_{l3}^{*}]$ and similarly, the upper bound $\alpha_{ub}^{*}$ can be obtained as $\alpha_{ub}^{*}=\min[\alpha_{u1}^{*}, \alpha_{u2}^{*}, \alpha_{u3}^{*}]$. 
\end{IEEEproof}

Global-optimal solution of $(J7)$ is given by Lemma $8$.
\begin{Lemma}
\textit{The global-optimal power control solution $\alpha_{sop}^{*}$ at which maximum SOP between users is minimized, is given as
\begin{equation}\label{optimalnew_a_asy_sop}
\!\alpha_{sop}^{*}\! \stackrel{\Delta}{=}\!\underset{\alpha \in \{\alpha_{1}^{*}, \alpha_{2}^{*}, \alpha_{3}^{*}\}}{\mathrm{argmin}}\! \max[s_{o1},s_{o2}],
\end{equation}
subject to the feasibility of candidate optimal points as $\alpha_{1}^{*}, \alpha_{2}^{*}, \alpha_{3}^{*}\in [\alpha_{lb}^{*}, \alpha_{ub}^{*}]$.}
\end{Lemma}
\begin{IEEEproof}
Since $(J7)$ is a nonconvex optimization problem due to the presence of nonconvex constraints $(C3)$ and $(C4)$, the candidate optimal points $(\alpha_{1}^{*}, \alpha_{2}^{*}, \alpha_{3}^{*})$ are obtained similar to the proof of Lemma $6$. In order to ensure users' rates necessity, we check if the obtained candidate optimal points satisfy POP constraint $(C2)$. Knowing $\alpha_{lb}^{*}$ and $\alpha_{ub}^{*}$ as obtained in the proof of Lemma $7$, feasibility of optimal points can be ensured. After investigating feasible optimal points, the global-optimal solution $\alpha_{sop}^{*}$ is obtained at that feasible optimal point for which maximum SOP between users is minimum. 
\end{IEEEproof}

Fig. $2$ presents the methodology to find optimal PA for pair outage constrained secrecy fairness maximization problem.

\subsection{Proposed Algorithm and Complexity Analysis}
\begin{algorithm}[!t]
{\small
\caption{\small Algorithm to find global-optimal PA $\alpha_{sop}^{*}$.}\label{Algo:AL1}
\begin{algorithmic}[1]
\Require $n$, $L_{p}$, $d_{1}$, $d_{2}$, $\xi$, $R_{1}^{th}$, $R_{2}^{th}$, $R_{s1}^{th}$, $R_{s2}^{th}$ and $\rho_{t}$  
\Ensure $\alpha_{sop}^{*}$ \newline
\textbf{(A) Finding potential candidates}
\State Obtain $\alpha_{1}^{*}$, $\alpha_{2}^{*}$, $\alpha_{3}^{*}$ for minimizing $s_{o1}$, $s_{o2}$, $s_{o1}=s_{o2}$, respectively, using golden section search algorithm
\newline
\textbf{(B) Check Feasibility}
\State Obtain $\alpha_{lb}^{*}$ and $\alpha_{ub}^{*}$ by solving the equation $p_{o}=\xi$ where $p_{o}$ is given in \eqref{pair_outage3}
\If{$\alpha_{lb}^{*} < \alpha_{1}^{*} < \alpha_{ub}^{*}$}
\State $\alpha_{1}^{*}$ is feasible optimal point
\Else
\State update $\alpha_{1}^{*} = \alpha_{lb}^{*} $
\EndIf
\If{$\alpha_{lb}^{*} < \alpha_{2}^{*} < \alpha_{ub}^{*}$}
\State $\alpha_{2}^{*}$ is feasible optimal point
\Else
\State update $\alpha_{2}^{*} = \alpha_{ub}^{*} $
\EndIf
\If{$\alpha_{lb}^{*} < \alpha_{3}^{*} < \alpha_{ub}^{*}$}
\State $\alpha_{3}^{*}$ is feasible optimal point
\Else
\State update $\alpha_{3}^{*}$ = infeasible point 
\EndIf
\newline
\textbf{(C) Global-optimal Solution}
\State Calculate $\alpha_{sop}^{*} \stackrel{\Delta}{=} \underset{\alpha \in \{\alpha_{1}^{*}, \alpha_{2}^{*}, \alpha_{3}^{*}\}}{\mathrm{argmin}} \max[s_{o1},s_{o2}]$ 
\end{algorithmic}
}
\end{algorithm}

Next, we present an algorithm that finds global-optimal PA which maximizes secrecy fairness between users while fulfilling their rate demands. For a given set of simulation parameters, $\alpha_{lb}^{*}$ and $\alpha_{ub}^{*}$ need to be calculated just once. Also, golden section search is applied to find three candidate optimal points of nonconvex optimization problem $(J7)$. Once they are calculated, $\alpha_{sop}^{*}$ is obtained by following the steps detailed in Algorithm 1. In order to investigate the complexity of proposed algorithm solving the optimization problem, we calculate total number of computations. The proposed algorithm is based on golden section search which terminates after $N$ iterations if $(\alpha_{ub}^{*}-\alpha_{lb}^{*})\times 0.618^N \leq \epsilon$. Hence, the total number of iterations in the proposed algorithm are $3N$, where $
N = 2\ln \Big(\frac{\alpha_{ub}^{*}-\alpha_{lb}^{*}}{\epsilon}\Big)$. 

\subsection{Asymptotic Power Control for min-max Secrecy Outage}
In above analysis, we solved the min-max SOP optimization problem numerically. In order to gain analytical insights, now we derive the asymptotic closed-form expression of global-optimal PA under high SNR region. Considering the users' rate requirements, problem can be stated as
\begin{align}\label{problem_formulation_qos_asy}
(J9): \underset{\alpha}{\text{minimize }} && \max[\hat s_{o1},\hat s_{o2}],&& \text{s.t.} && (C1), (C2). 
\end{align}

Following $\hat x_{c} \stackrel{\Delta}{=} \max[\hat s_{o1},\hat s_{o2}]$, an equivalent formulation of $(J9)$ can be given as 
\begin{align}\label{problem_formulation1_qos_asy}
(J10):  \underset{\alpha, \hat x_{c}}{\text{minimize}} \quad \hat x_{c}, \quad
\text{s.t.} \quad (C1), (C2), \nonumber
\\(C5): \hat s_{o1}\leq \hat x_{c}, (C6): \hat s_{o2}\leq \hat x_{c},
\end{align}
where $(C5)$ and $(C6)$ appears from the definition of max$[\cdot]$. 

We first solve optimization problem without considering QoS constraint, which can be given as 
\begin{align}\label{problem_formulation1_no_qos_asy}
&(J11):  \underset{\alpha, \hat x_{c}}{\text{minimize }} && \hat x_{c},&& 
\text{ s.t. } &&  (C1), (C5), (C6). 
\end{align}

Global-optimal solution of $(J11)$ is provided by Lemma $9$.
\begin{Lemma} \label{lemma_2}
\textit{The global-optimal solution $\hat \alpha$ of the optimization problem $(J11)$ for maximizing the secrecy fairness is given by
\begin{equation}\label{optimal_asy_a}
\dot \alpha_{sop} \stackrel{\Delta}{=} \underset{\alpha \in \{\hat \alpha_{1}, \hat \alpha_{2}, \hat \alpha_{3}\}}{\mathrm{argmin}}\max[\hat s_{o1},\hat s_{o2}],
\end{equation}
where $\hat \alpha_{1}, \hat \alpha_{2}$, respectively, have been obtained by minimizing $\hat s_{o1}, \hat s_{o2}$ (cf. \ref{individual_sop_asymp}), and $\hat \alpha_{3}$ is obtained by solving $\hat s_{o1}=\hat s_{o2}$.}
\end{Lemma}
\begin{IEEEproof}
We associate Lagrange multipliers $\mu_{1}$ with $(C5)$, $\mu_{2}$ with $(C6)$, and keep the boundary constraint $(C1)$ implicit. Then, Lagrangian function $\mathcal{\hat L}$ can be written as
\begin{equation}\label{lagragian_function}
\mathcal{\hat L} \stackrel{\Delta}{=} \hat x_{c} + \mu_{1}[\hat s_{o1} - \hat x_{c}] + \mu_{2}[\hat s_{o2} - \hat x_{c}].
\end{equation}

The corresponding KKT conditions are given by constraints $(C5)$ and $(C6)$. Dual feasibility conditions are obtained as $\mu_{1}\geq 0$ and $\mu_{2}\geq 0$ using \eqref{problem_formulation1_no_qos_asy} and \eqref{lagragian_function}. The subgradient conditions are given as $
\frac{\mathrm{d}\mathcal{\hat L}}{\mathrm{d}\hat x_{c}}  =  1-\mu_{1}-\mu_{2}= 0,$
and $\frac{\mathrm{d}\mathcal{\hat L}}{\mathrm{d}\alpha}  =  \mu_{1} \frac{\mathrm{d}\hat s_{o1}}{\mathrm{d}\alpha}  + \mu_{2}\frac{\mathrm{d}\hat s_{o2}}{\mathrm{d}\alpha} = 0$. The complementary slackness conditions are obtained as
$\mu_{1} [\hat s_{o1} - \hat x_{c}] = 0$ and $\mu_{2} [\hat s_{o2} - \hat x_{c}] = 0$. Here also, three cases exist. \emph{Case 1: $\mu_{1} > 0$ and $\mu_{2}= 0$}, implies $\frac{\mathrm{d}\hat s_{o1}}{\mathrm{d}\alpha}=0$ and results $\alpha = \hat \alpha_{1}$ from $\hat s_{o1}$ minimization for U$1$ \eqref{optimal_a_asy_sop1}. \emph{Case 2: $\mu_{2} > 0$ and $\mu_{1}= 0$}, implies $\frac{\mathrm{d}\hat s_{o2}}{\mathrm{d}\alpha}=0$ and provides the same solution as the optimal solution of $\hat s_{o2}$ minimization, i.e., $\alpha = \hat \alpha_{2}$ from \eqref{optimal_a_asy_sop2}. \emph{Case 3: $\mu_{1} > 0$ and $\mu_{2} > 0$}, implies $\hat s_{o1} = \hat s_{o2}$ and it gives $\hat \alpha_{3}$ which is given as
\begin{equation}
\hat \alpha_{3} = \frac{ \gamma_{12}\Pi_{2}\lambda_{1} + (\rho_{t}\beta_{21}-\gamma_{21}\Pi_{1})\lambda_{2} }{\gamma_{12}\lambda_{1} + \gamma_{21}\lambda_{2}}.
\end{equation}

Observing that the minimization problem has three optimal points, i.e., $\hat \alpha_{1}$, $\hat \alpha_{2}$, and $\hat \alpha_{3}$, the global optimal solution $\dot \alpha_{sop}$ is obtained as the point where maximum SOP is minimum.
\end{IEEEproof}

Next, while taking the QoS constraint into account, the global-optimal solution of  $(J10)$ is given by Lemma $10$.
\begin{Lemma}
\textit{Global-optimal solution $\hat \alpha_{sop}$ of optimization problem $(J10)$ for minimizing the maximum SOP is given by
\begin{equation}\label{optimal_asy_a_asy}
\hat \alpha_{sop} \stackrel{\Delta}{=} \underset{\alpha \in \{\hat \alpha_{1}, \hat \alpha_{2}, \hat \alpha_{3}\}}{\mathrm{argmin}}\max[\hat s_{o1},\hat s_{o2}],
\end{equation}
subject to the feasibility of candidate optimal points as $\hat \alpha_{1}, \hat \alpha_{2}, \hat \alpha_{3} \in [\alpha_{lb}^{*}, \alpha_{ub}^{*}]$.}
\end{Lemma}
\begin{IEEEproof}
Similar to proof of Lemma $9$, we first find the candidate optimal points $\hat \alpha_{1}, \hat \alpha_{2},$ and $\hat \alpha_{3}$. Next, the feasibility of these points is investigated between $\alpha_{lb}^{*}$ and $\alpha_{ub}^{*}$. Finally, the global optimal point is obtained as the feasible optimal point where maximum SOP is minimized. 
\end{IEEEproof}

\section{Numerical Investigations}
\begin{figure*}[!t]	
	\begin{minipage}{.48\textwidth}\color{black}	
		\centering\includegraphics[height=2.3in,width=3.6in]{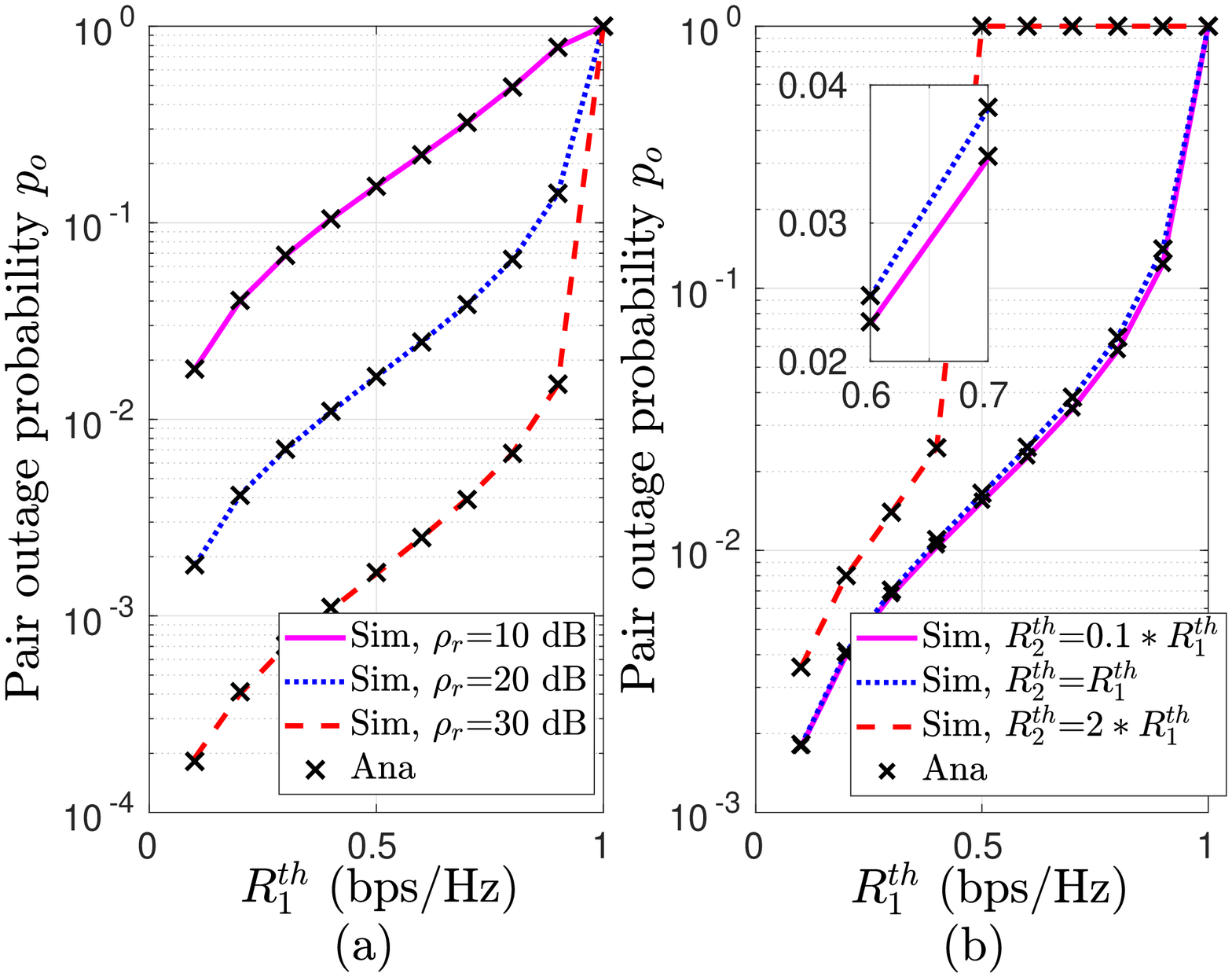}
		\vspace{-5mm}
		\caption{Validation of pair outage probability $p_{o}$ with users' threshold rates, (a) $R_{2}^{th}=R_{1}^{th}$, and (b) $\rho_{r}=20$ dB.}
		\label{validationPOP1}\color{black} 
	\end{minipage}\quad\
	\begin{minipage}{.48\textwidth}	
		\centering\includegraphics[height=2.3in,width=3.6in]{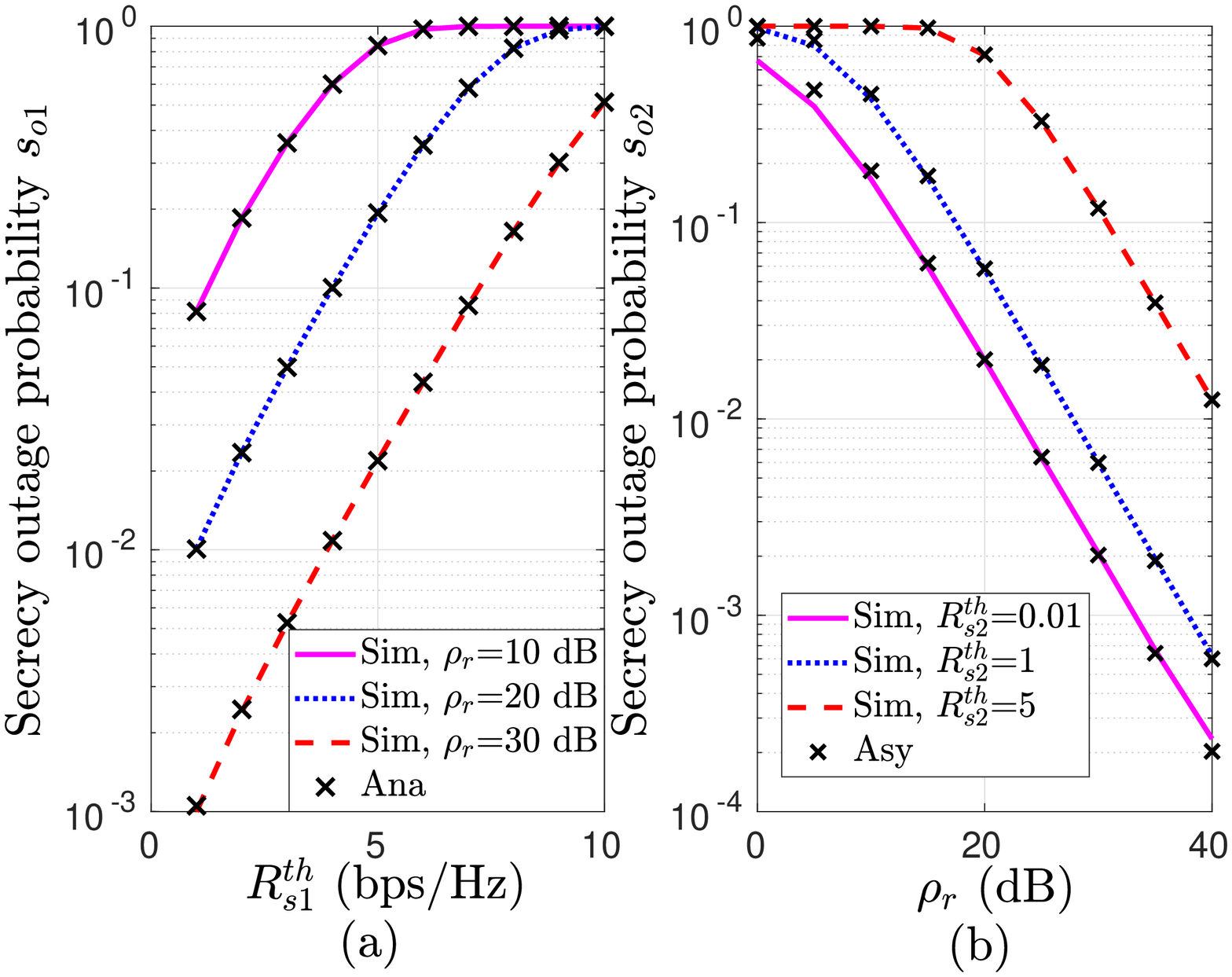}
		\vspace{-5mm}
		\caption{Validation of secrecy outage probability, (a) $s_{o1}$ for U$1$, and (b) $s_{o2}$ for U$2$.}
		\label{f1SOP1} 
	\end{minipage} \vspace{-2mm} 
\end{figure*}

\begin{figure*}[!t]	
	\begin{minipage}{.48\textwidth}\color{black}	
		\centering\includegraphics[height=2.3in,width=3.6in]{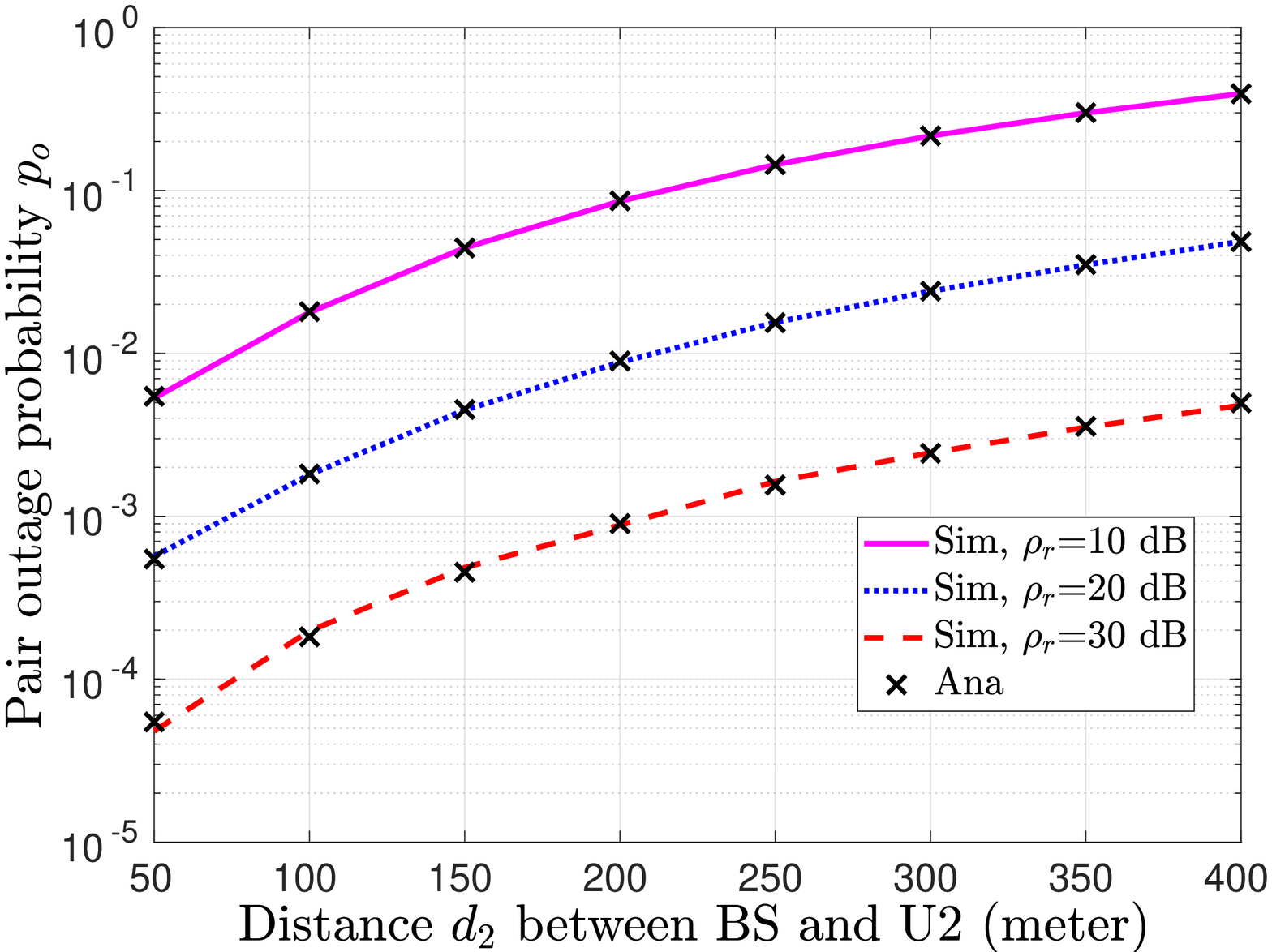}
		\vspace{-5mm}
		\caption{Variation of pair outage probability $p_{o}$ versus U$2$'s distance $d_{2}$ with $R_{2}^{th}=R_{1}^{th}=0.1$.}
		\label{distancePOP}\color{black} 
	\end{minipage}\quad\;
	\begin{minipage}{.48\textwidth}	
		\centering\includegraphics[height=2.3in,width=3.6in]{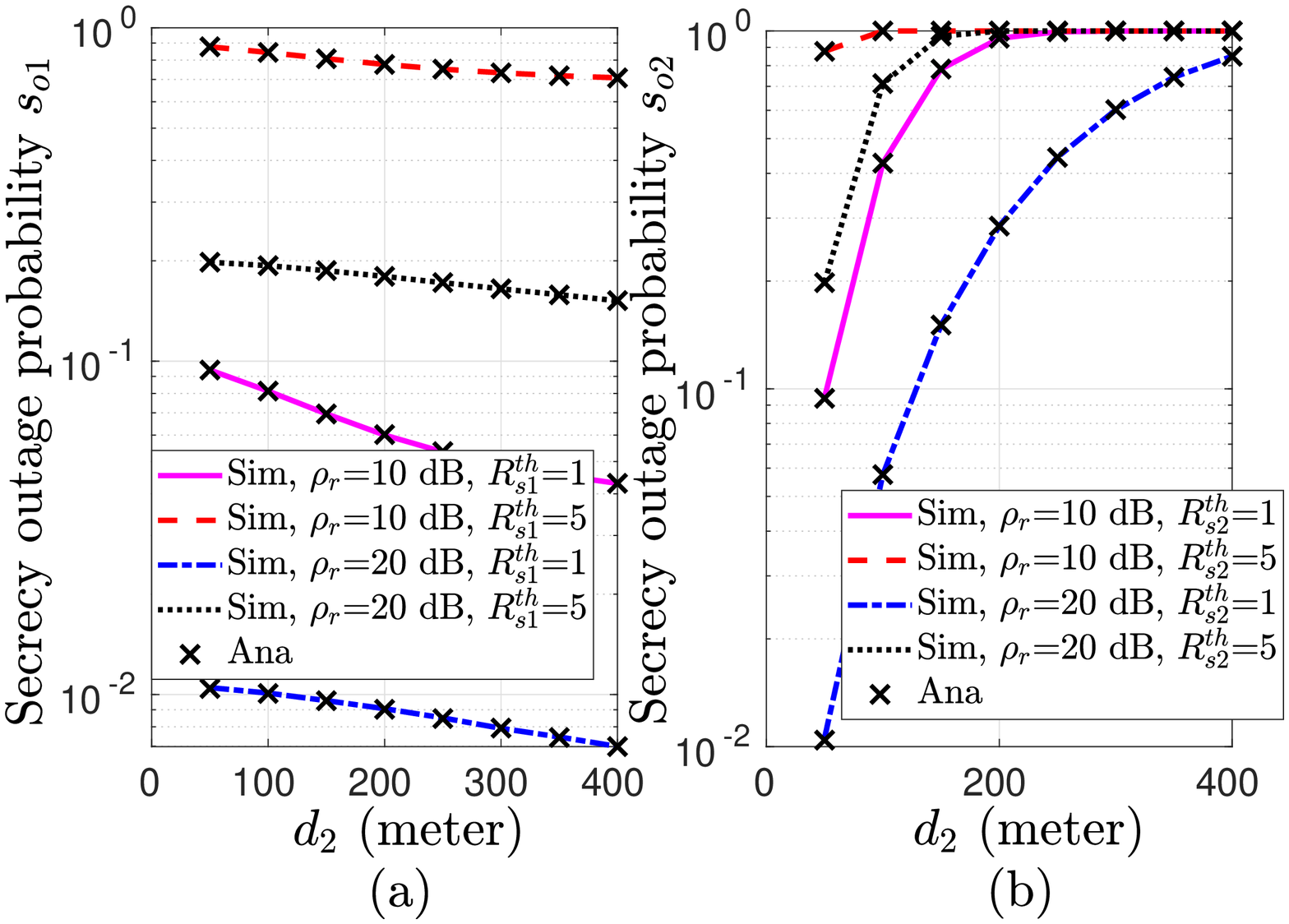}
		\vspace{-5mm}
		\caption{Variation of secrecy outage probability versus U$2$'s distance $d_{2}$, (a) $s_{o1}$ for U$1$, and (b) $s_{o2}$ for U$2$.}
		\label{distance_variation1}
	\end{minipage} \vspace{-2mm} 
\end{figure*}

Next, numerical results are presented to assess the performance of proposed protocol. We present simulation setup, and discuss the system performance under various system settings. Specifically, we validate accuracy of analytical results, provide insights on effect of users' distance and optimal PAs, and identify the performance gains achieved by global-optimal PA. Finally, tradeoff between QoS parameter POP and secrecy performance SOP is presented.

\subsection{Simulation Setup}
Downlink of a NOMA system is considered, where BS communicates with two untrusted users. For the sake of simulation study, near user distance $d_{1}$ and far user distance $d_{2}$ from BS are taken as $50$ meter and $100$ meter, respectively. $L_{p}$ and $n$ are taken as $1$ and $2.5$, respectively. Noise signal at both users follow Gaussian distribution with a noise power of $-60$ dBm. Small scale fading affecting both links is assumed to have exponential distribution with mean value $1$ \cite{basepaper}. Residual interference terms $\beta_{21}$ and $\beta_{12}$ are assumed to be equal and $-30$ dBm \cite{8370069}. Simulation results are averaged over $10^6$ randomly generated channel realizations for both U$1$ and U$2$. The correctness of results is observed using the fact that root mean square error (RMSE) in approximating the derived results should be less than $0.0003$ \cite{7073612}. For golden section search, $\epsilon=0.01$. $\rho_{r}$ is considered as the received SNR. POP and SOP are considered as performance metrics to evaluate the performance of system.

\subsection{Validation of Analysis}
We first validate analytical expressions of $p_{o}$ in Fig. \ref{validationPOP1}, and $s_{o1}$ and $s_{o2}$ in  
Fig. \ref{f1SOP1}. For these results, $\alpha$ is considered as $0.5$. A close match between the simulation and analytical results validates the analysis of $p_{o}$, $s_{o1}$, and $s_{o2}$ with a RMSE of the order of $10^{-4}$. Accuracy of asymptotic results, marked as `Asy', is also verified with simulation results in Fig. \ref{f1SOP1}(b). We observe that $p_{o}$ increases with increase in the threshold rates $R_{1}^{th}$ or $R_{2}^{th}$. Similarly, increasing target secrecy rates $R_{s1}^{th}$ and $R_{s2}^{th}$ increase $s_{o1}$ and $s_{o2}$, respectively. Following the fact that the outage occurs when the users' maximum achievable rate is below a threshold rate, it is natural that increasing threshold rates and target secrecy rates at users, respectively, increases POP and SOP. 

From the results it is also observed that threshold rate pair ($R_{1}^{th}, R_{2}^{th}$) with lesser value of $R_{2}^{th}$ compared to $R_{1}^{th}$ results a performance improvement in POP. The reason is that the achievable information rate at U$2$ is always less in comparison to U$1$ due to the poorer channel condition of U$2$. On the other side, outage happens when the rate falls below a threshold rate. Therefore, in such a case, if more value of $R_{2}^{th}$ is considered, outage probability at U$2$ will be increased, which will further increase POP. Therefore, it can be concluded that the threshold rate pair ($R_{1}^{th}, R_{2}^{th}$) with $R_{2}^{th} < R_{1}^{th}$ has more practical significance to obtain better pair outage performance.

Further, we observe that increasing $\rho_{r}$ decreases $p_{o}$, $s_{o1}$ and $s_{o2}$. With increasing SNR, achievable data rates and secrecy rates at users increase, hence, for a given threshold rate and target secrecy rate, POP and SOP decreases, respectively. Noting the above observations, we can conclude that the performance of a system is highly dependent on system parameters.

\begin{figure*}[!t]	
	\begin{minipage}{.48\textwidth}\color{black}	
		\centering\includegraphics[height=2.3in,width=3.6in]{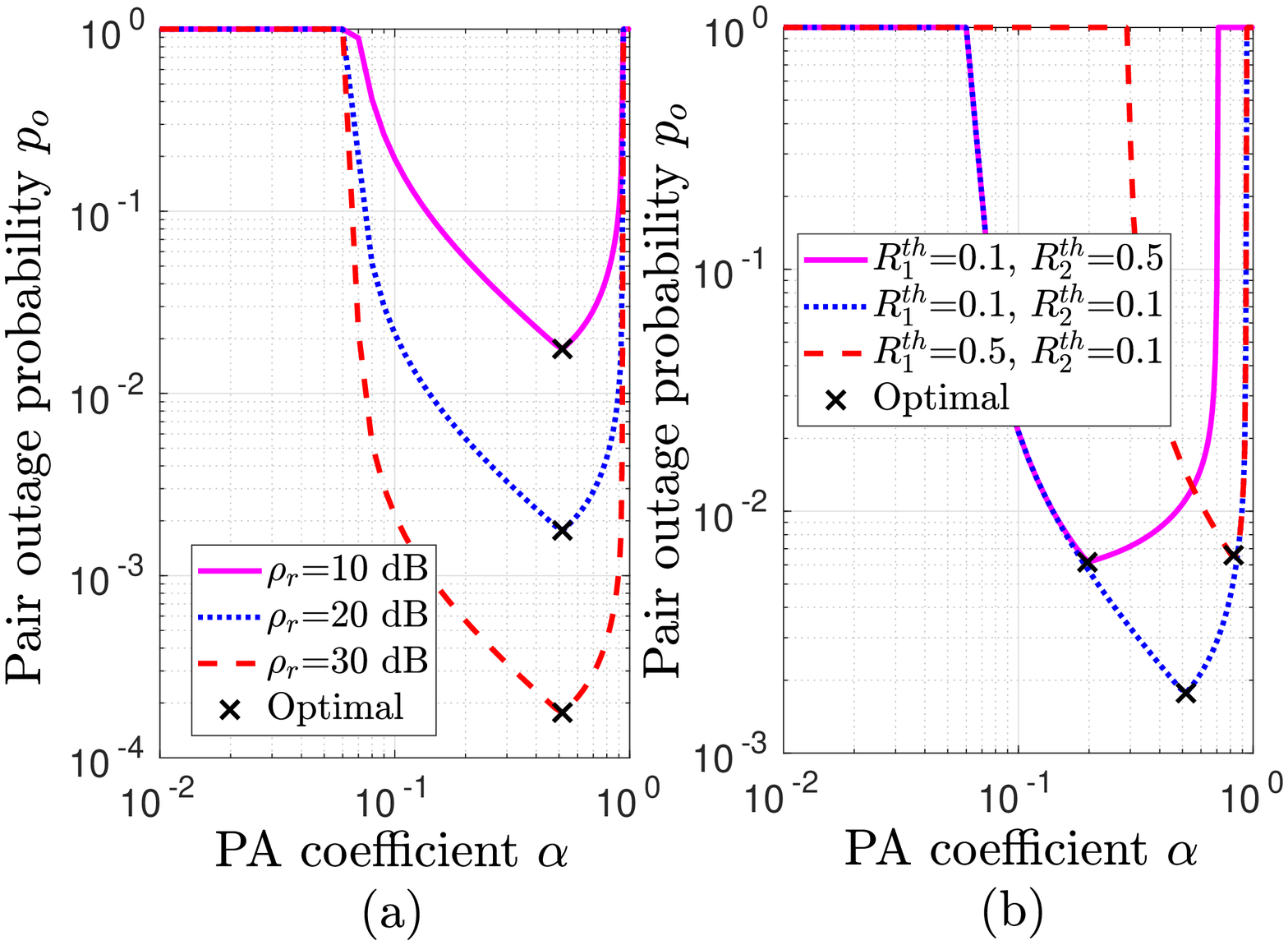}
		\vspace{-5mm}
		\caption{Pseudo convexity of POP through $p_{o}$ versus $\alpha$ analysis, (a) $R_{2}^{th}=R_{1}^{th}=0.1$, and (b) $\rho_{r}=20$ dB.}
		\label{optimal_outagePOP1}\color{black} 
	\end{minipage}\quad\;
	\begin{minipage}{.48\textwidth}	
		\centering\includegraphics[height=2.3in,width=3.6in]{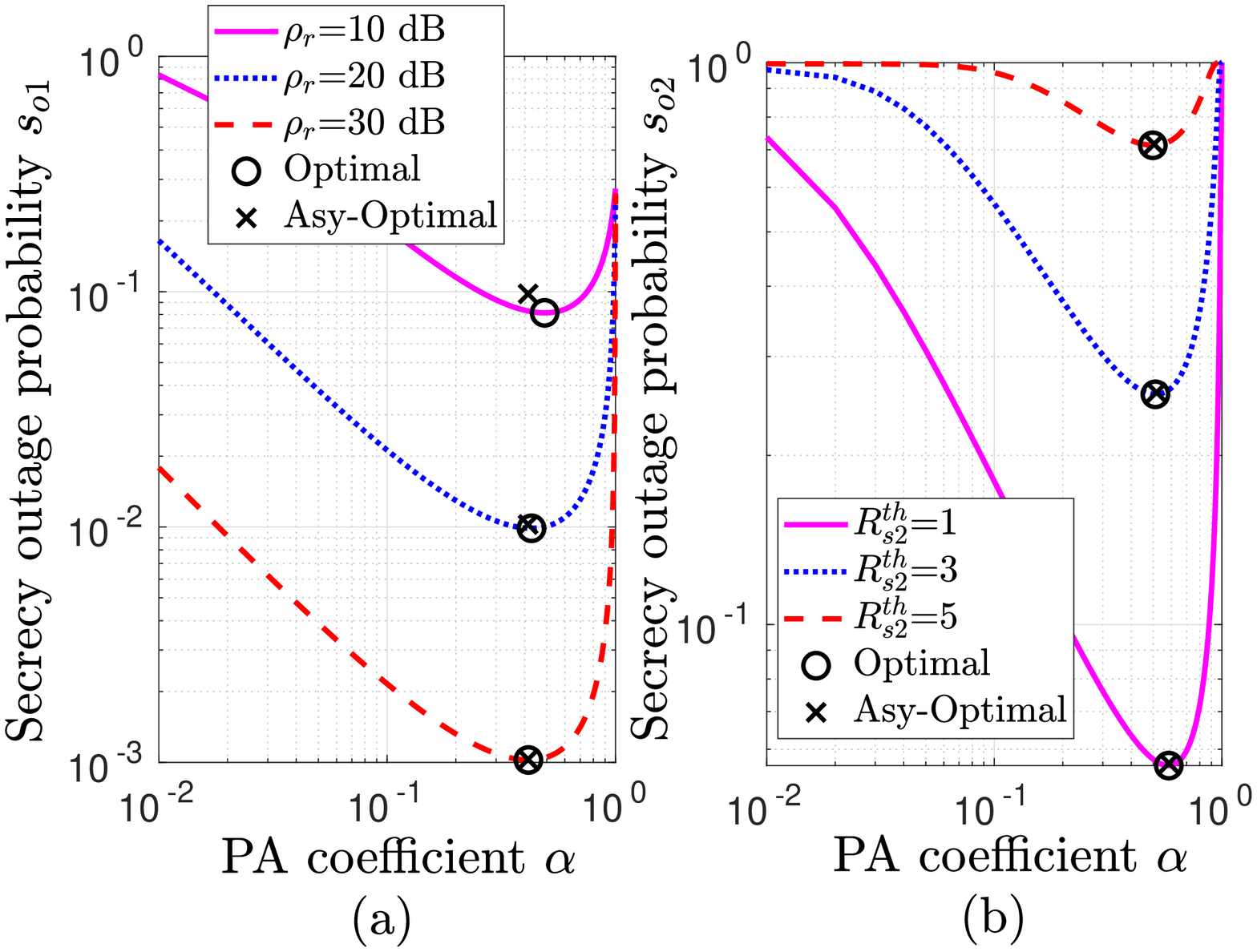}
		\vspace{-5mm}
		\caption{Pseudo convexity of SOP through variation of SOP with $\alpha$, (a) $s_{o1}$ at $R_{s1}^{th}=1$, and (b) $s_{o2}$ at $\rho_{r}=20$ dB.}
		\label{optimal_outage1}
	\end{minipage} \vspace{-2mm} 
\end{figure*}

\begin{figure*}[!t]	
	\begin{minipage}{.48\textwidth}\color{black}	
		\centering\includegraphics[height=2.3in,width=3.6in]{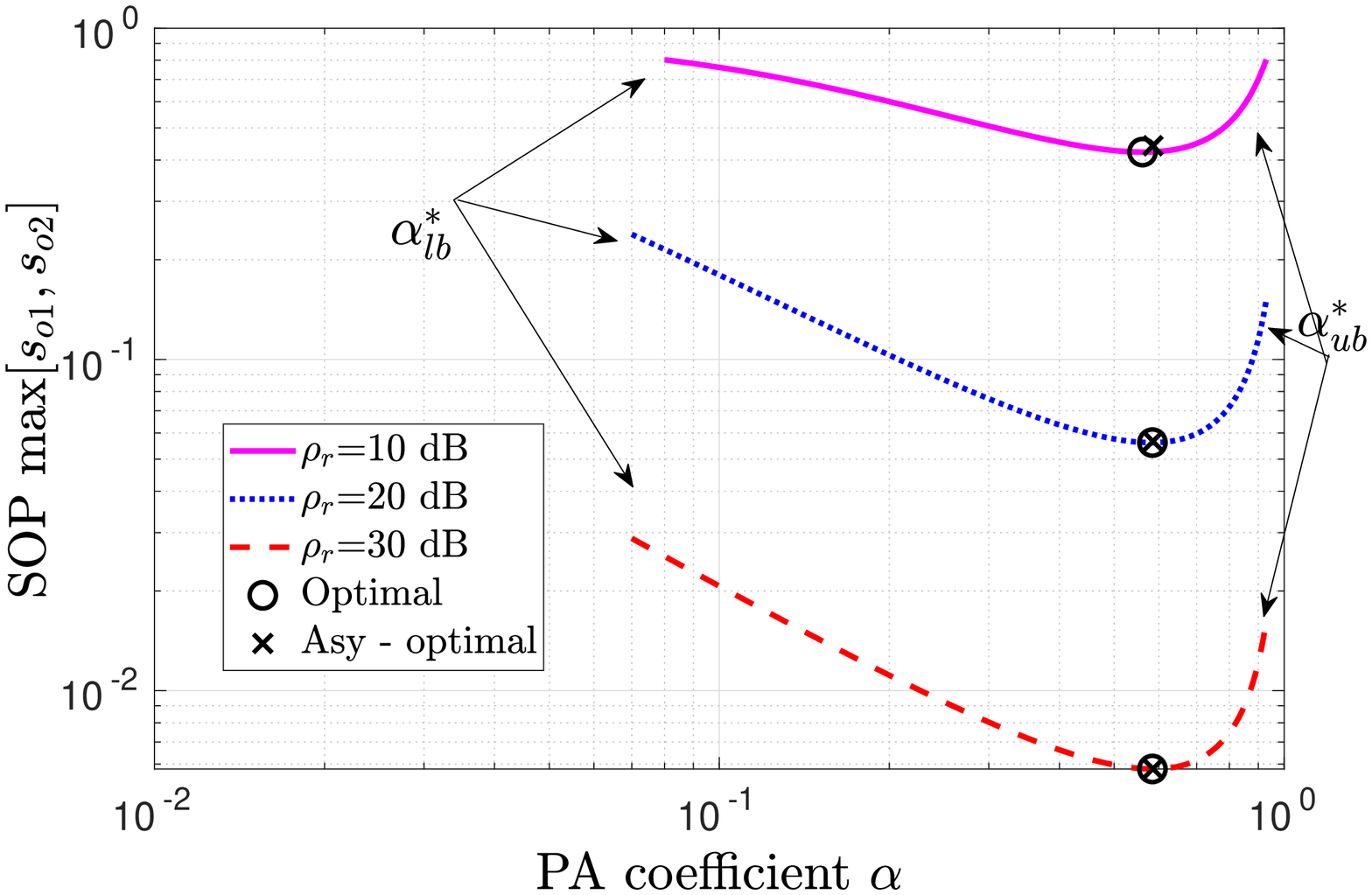}
		\vspace{-5mm}
		\caption{Optimal SOP $\max[s_{o1},s_{o2}]$ and PA coefficient $\alpha$ analysis with $R_{s2}^{th}=R_{s1}^{th}=1$, $R_{2}^{th}=R_{1}^{th}=0.1$, and POP constraint $\xi=0.5$.}
		\label{optimal_max_SOP}\color{black} 
	\end{minipage}\quad\;
		\begin{minipage}{.48\textwidth}	
		\centering\includegraphics[height=2.3in,width=3.6in]{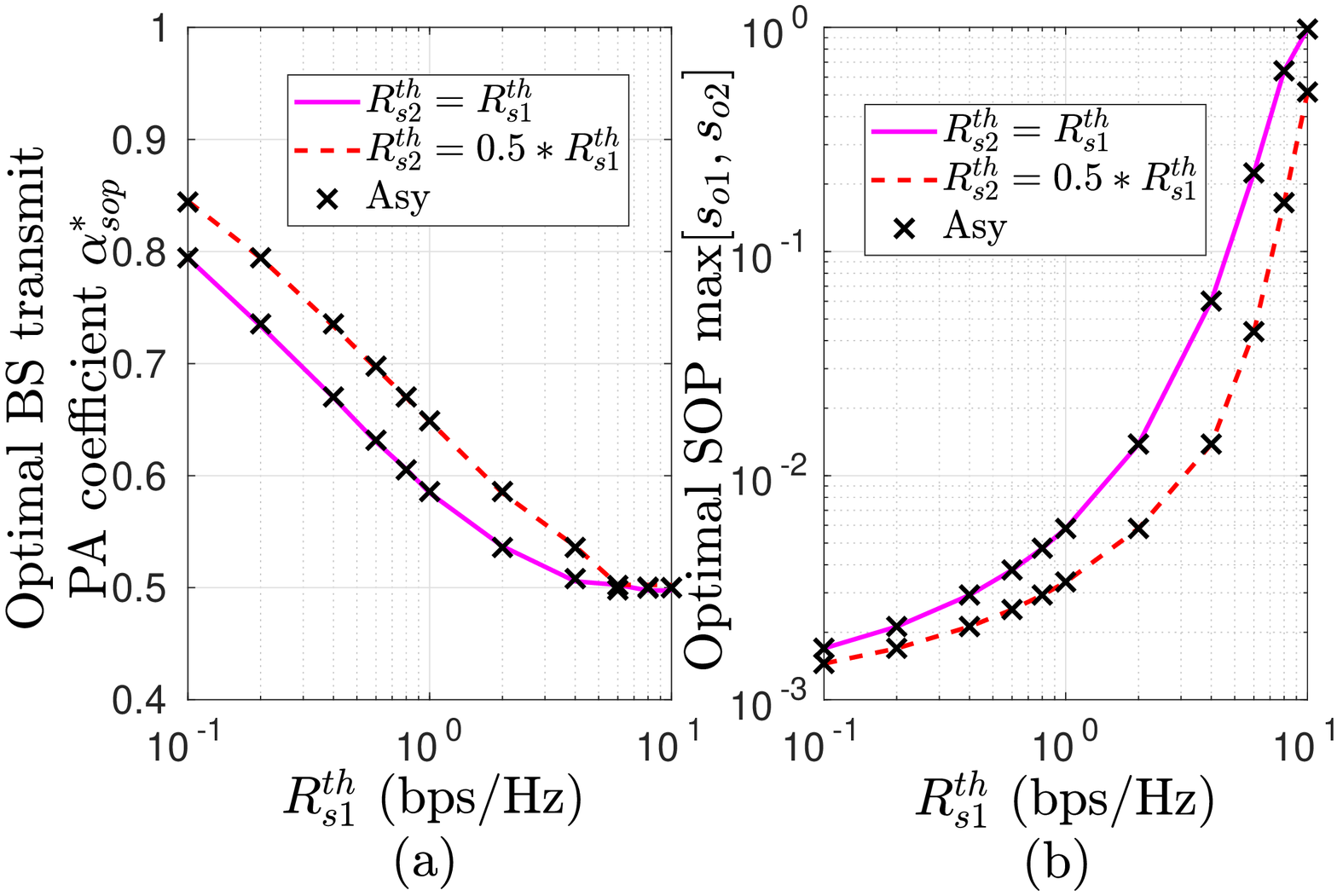}
		\vspace{-5mm}
		\caption{(a) Optimal PA with target secrecy rate $R_{s1}^{th}$, and (b) secrecy fairness $\min \max[s_{o1},s_{o2}]$ with $R_{s1}^{th}$, $\rho_{r}=30$ dB, $R_{2}^{th}=R_{1}^{th}=0.1$, $\xi=0.5$.}
		\label{a_comp}
	\end{minipage} \vspace{-2mm} 
\end{figure*}

\subsection{Impact of Relative Distance between Users}
Fixing $d_{1}$ as $50$ meter, now we observe the effect of variation of $d_{2}$ from BS on achievable POP and SOP. In Fig. \ref{distancePOP}, increasing $d_{2}$ implies decrease in achievable data rate and hence, increasing the outage probability of U$2$ which results in an increase in POP. On the contrary, the results presented in  Fig. \ref{distance_variation1}(a), show that $s_{o1}$ monotonically decreases with the increase in $d_{2}$. The reason is, with increasing distance $d_{2}$, a decrease in achievable information rate at U$2$ improves secrecy rate at U$1$, and hence, SOP at U$1$ decreases. Also, decrease in information rate at U$2$ implies decrease in secrecy rate at U$2$ which increases SOP for U$2$ as shown in Fig. \ref{distance_variation1}(b). Note that increasing the distance between BS to U$2$ has an opposite effect on $p_{o}$ and $s_{o1}$. Similarly, Fig. \ref{distance_variation1}(a) and Fig. \ref{distance_variation1}(b) also show a contradicting effect of $d_{2}$ on $s_{o1}$ and $s_{o2}$. Hence, it can be summarized that since POP and SOPs for both the users are important system parameters for reliable and secure NOMA communication and fulfilling them together is highly dependent on the relative distance between BS and users.

\begin{figure*}[!t]	
	\begin{minipage}{.48\textwidth}\color{black}	
		\centering\includegraphics[height=2.3in,width=3.6in]{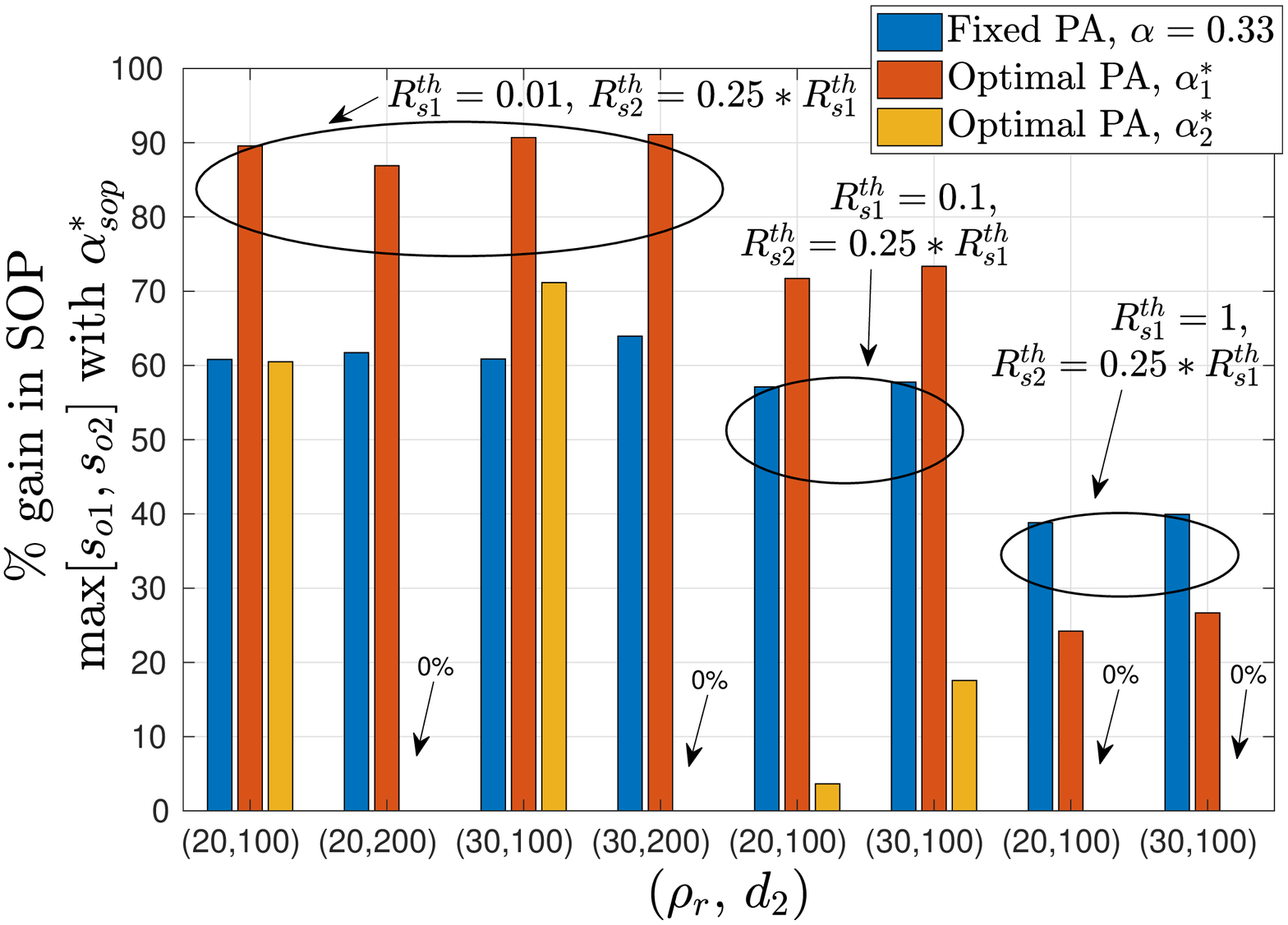}
		\vspace{-5mm}
		\caption{Performance comparison of global-optimal PA, $\alpha_{sop}^{*}$, with fixed PA $\alpha=0.33$, and individual optimal PA $\alpha_{1}^{*}$ and $\alpha_{2}^{*}$, $R_{2}^{th}=R_{1}^{th}=0.1$, $\xi=0.5$.}
		\label{performance_comp}\color{black} 
	\end{minipage}\quad\;
	\begin{minipage}{.48\textwidth}	
		\centering\includegraphics[height=2.3in,width=3.6in]{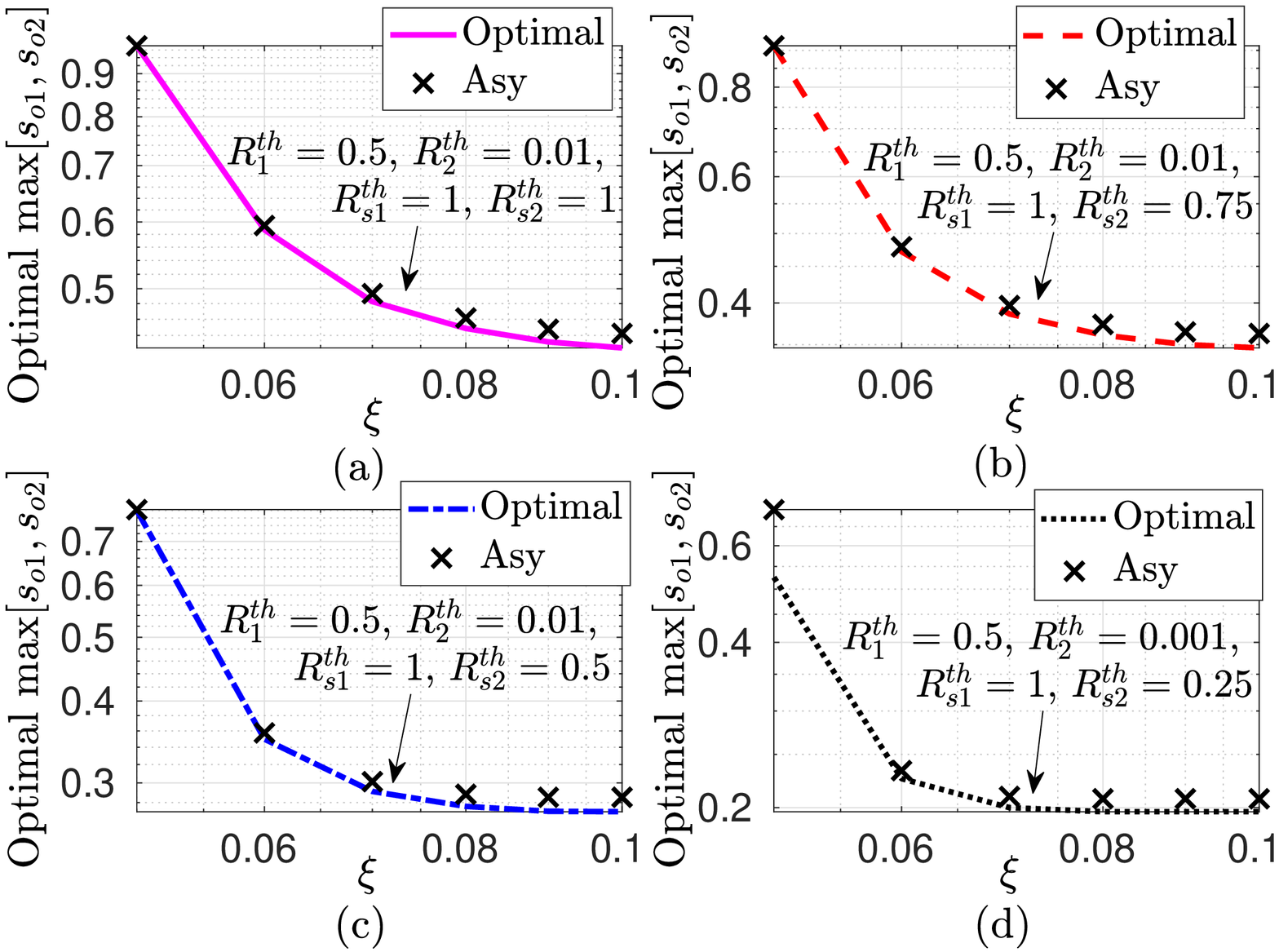}
		\vspace{-5mm}
		\caption{Tradeoff between optimal SOP $\max[s_{o1},s_{o2}]$ and POP threshold $\xi$ at SNR $\rho_{r}=10 $ dB for different secrecy rate targets.}
\label{tradeoff}
	\end{minipage} \vspace{-2mm} 
\end{figure*}

\subsection{Insights on optimality}
Now we investigate optimal pair outage and secrecy outage performance by presenting the numerical proof of the generalized-convexity of POP and SOP with respect to $\alpha$. Fig. \ref{optimal_outagePOP1}(a) and Fig. \ref{optimal_outagePOP1}(b) depict POP performance with $\alpha$ for different values of SNRs and threshold rates, respectively. The pseudoconvex nature of POP can be easily observed. It confirms the unique solution of PA that provides minimum POP. Analytical results have also been plotted as a validation to the analysis. Similarly, Fig. \ref{optimal_outage1}(a) and Fig. \ref{optimal_outage1}(b), respectively, validate pseudoconvex nature of $s_{o1}$ and $s_{o2}$ with $\alpha$ for different SNRs and threshold rates. The numerical optimal solution obtained using golden section algorithm has also been indicated. We observe that numerical results match with asymptotic results, which confirms the validity of asymptotic analysis. Further, to demonstrate secrecy fairness maximization under QoS constraint, Fig. \ref{optimal_max_SOP} presents $\max[s_{o1}$,$s_{o2}]$ with $\alpha$. $\alpha_{lb}^{*}$ and $\alpha_{ub}^{*}$ are shown as lower and upper bounds on $\alpha$, respectively, for an example case of $\xi=0.5$. A close match between numerical and asymptotically optimal results can be observed at $\rho_{r} \geq 20$ dB, which validates high SNR analysis. From the observations,  it is worth noting that PA $\alpha$ decides allocated powers to users, which effects POP and SOPs. Hence, an appropriate PA for given system parameters can help to obtain an optimal secure and reliable communication system.

Next, Fig. \ref{a_comp}(a) shows global-optimal $\alpha_{sop}^{*}$ that minimizes the maximum SOP between users as a function of target secrecy rates $R_{s1}^{th}$ for various values of $R_{s2}^{th}$. Results show that for each target secrecy rate pair ($R_{s1}^{th}$, $R_{s2}^{th}$), there is one and only one $\alpha$ such that $\alpha P_{t}$ and $(1-\alpha)P_{t}$ are the optimal powers for U$1$ and U$2$, respectively. On increasing $R_{s1}^{th}$, the optimal value of $\alpha$ decreases, whereas the optimal SOP obtained from min-max optimization problem, presented in Fig. \ref{a_comp}(b) increases. Further, lower value of $R_{s2}^{th}$ in comparison to $R_{s1}^{th}$ provides improvement in SOP which results in  a higher value of $\alpha$.
This is because the achievable data rate at U$2$ is less compared to the achievable data rate at U$1$ because of the weaker channel condition of U$2$, due to which the achievable secrecy rate at  U$2$ decreases. Note that according to the definition of secrecy outage, outage happens when the rate falls below a threshold rate. Therefore, it is natural if more value of $R_{s2}^{th}$ is considered, SOP for U$2$ will be increased. Therefore, we can conclude that the threshold secrecy rate pair ($R_{s1}^{th}, R_{s2}^{th}$) with $R_{s2}^{th} < R_{s1}^{th}$ is a better choice to improve system performance.

\subsection{Performance Comparison}
To demonstrate the performance gain achieved by the proposed algorithm for secrecy fairness maximization, 
Fig. \ref{performance_comp} presents the performance comparison of global-optimal PA $\alpha_{sop}^{*}$ with fixed PA $\alpha=0.33$, and individual optimal PAs $\alpha_{1}^{*}$ and $\alpha_{2}^{*}$ obtained by  minimizing $s_{o1}$ and $s_{o2}$, respectively. Percentage gain depicts that $\alpha_{sop}^{*}$ achieves best SOP performance, because it ensures secrecy fairness between U$1$ and U$2$. The average percentage improvement by $\alpha_{sop}^{*}$ over fixed PA, individual optimal PAs $\alpha_{1}^{*}$ and $\alpha_{2}^{*}$ are respectively around $55.12\%$, $69.30\%$ and $19.11\%$. Thus, it can be concluded that the global optimal PA allocated to users plays an important role in providing the best system performance.

\subsection{Tradeoff}
Secrecy performance and users' data rates demands both are important system performance parameters for secure and reliable communication. However, fulfilling them together is a challenge because when QoS is considered, system sacrifices secrecy rate to improve QoS demands. This interesting tradeoff between minimized SOP and POP, as shown in Fig. \ref{tradeoff}, emphasizes the effect of QoS demands on optimal SOP performance. Decrease in SOP with an increase in POP highlights that higher rate demands lead to poor secrecy performance.

\section{Concluding Remarks}
This paper has proposed a novel decoding order that is capable of providing positive secrecy rate for both near and far users in a two-user untrusted NOMA system. To analyze secrecy performance, SOP for both users are derived, and closed-form expressions are given for the high SNR regime. Individual PA optimization to minimize SOPs and asymptotic optimal solutions have also been presented. In order to ensure users’ QoS demands for reliable communication over all the links, POP has been derived as a QoS measure, and optimal PA minimizing POP has also been obtained. Further, with an intention to provide secrecy fairness between users while satisfying their predefined QoS demands, optimization of PA to minimize the maximum SOP between users is also investigated. Numerical results are provided to verify the correctness of analytical expressions, provide insights on generalized-convexity of POP and SOP, highlight significant performance gains achieved by global-optimal PA, and describe the tradeoff between different system performance requirements. While this work has considered a two user NOMA system, the work can be extended to the study of secrecy in a multi-user scenario and considering other variants like code-domain NOMA. Furthermore, for an untrusted users’ scenario, it would be interesting to compare the performance of a NOMA enabled system with jammer assisted orthogonal multiple access. 

\begin{appendices}
\section{Proof of Theorem 2}
Having obtained feasible decoding orders for NOMA system with the positive secrecy rate consideration at all users, we compare these orders in terms of users' secrecy rates to investigate the optimal decoding order. The key idea here is, to find out that one decoding order which gives best secrecy rates at users. In this context, we have observed that, for each decoding order, secrecy rates $R_{s1}$ and $R_{s2}$ at U$1$ and U$2$, respectively, are in the form of $\log_{2}\frac{1+(A/B)}{1+(C/D)}$. Therefore, on comparing two decoding orders, we can easily show that a given secrecy rate is higher than its counterpart if either of $A$ or $D$ is higher, or either of $B$ or $C$ is lower, with other respective parameters being same. Thus, we can find out the optimal decoding order having maximum secrecy rate.

Considering Theorem $1$, there exists three feasible decoding orders: $(2,1)$, $(1,2)$, and $(1,1)$. We first compare $(2,1)$ and $(1,2)$. In $(2,1)$, $B=\beta_{21}+\frac{1}{\rho_{t}}$ and $D=(1-\alpha)|h_{2}|^{2}+\frac{1}{\rho_{t}}$ for $R_{s1}$, and, $B=\beta_{12}+\frac{1}{\rho_{t}}$ and $D=\alpha|h_{1}|^{2}+\frac{1}{\rho_{t}}$ for $R_{s2}$. Whereas, in $(1,2)$, $B=(1-\alpha)|h_{1}|^{2}+\frac{1}{\rho_{t}}$ and $D=\beta_{22}+\frac{1}{\rho_{t}}$ for $R_{s1}$, and, $B=\alpha|h_{2}|^{2}+\frac{1}{\rho_{t}}$ and $D=\beta_{11}+\frac{1}{\rho_{t}}$ for $R_{s2}$. The remaining parameters, i.e., $A$ and $C$ are the same for the two considered orders for $R_{s1}$. The same holds true for $R_{s2}$. Observing $R_{s1}$ in $(2,1)$  and $(1,2)$, we note that $B$ in $(2,1)$ is lower in comparison to $B$ in $(1,2)$, i.e., $\beta_{21}<(1-\alpha)|h_{1}|^{2}$. Also, $D$ in $(2,1)$ is higher  compared to $D$ in $(1,2)$, i.e., $(1-\alpha)|h_{2}|^{2}>\beta_{22}$. This is because $\beta_{ij}$ is the residual interefence from imperfectly decoded U$i$ by U$j$. Similarly, for $R_{s2}$ also, $B$ is lower and $D$ is higher for $(2,1)$ in comparison to $B$ and $D$, respectively, of $(1,2)$. Therefore, it can be concluded that secrecy rate in $(2,1)$ is higher than $(1,2)$ for both U$1$ and U$2$. In the similar manner, we compare $(2,1)$ and $(1,1)$. In $(1,1)$, $B=(1-\alpha)|h_{1}|^{2}+\frac{1}{\rho_{t}}$ for $R_{s1}$ and $D=\beta_{11}+\frac{1}{\rho_{t}}$ for $R_{s2}$. Here also, for $R_{s1}$ in $(2,1)$ and $(1,1)$, $B$ is lower in $(2,1)$ compared to the $B$ in $(1,1)$, and, for $R_{s2}$, $D$ is higher in $(2,1)$ in comparison to the $D$ in $(1,1)$. Thus, $(2,1)$ will  provide better secrecy rate than $(1,1)$ also, at both users. 

Thus, we conclude that out of the three feasible decoding orders that can provide positive secrecy rate at both users, $(2,1)$ is a decoding order that ensures maximum secrecy rate at both the users as compared to other two decoding orders. Therefore, $(2,1)$ is considered as an optimal decoding order regardless of any PA, which is followed in analysis further.

\section{Proof of Lemma 4 }
Note that $\hat \alpha_{1}$ is obtained by minimizing $\hat s_{o1}$ \eqref{sop1_asy}. The second-order derivative of $\hat s_{o1}$ with respect to $\alpha$ does not imply monotonicity. By setting $\frac{\mathrm{d}\hat s_{o1}}{\mathrm{d}\alpha}=0$, we obtain $\hat \alpha_{1} = - (\Pi_{1}-1) \pm \sqrt{(\Pi_{1}(\Pi_{1} -1)}$. We observe that $\hat \alpha_{1}  = - (\Pi_{1}-1) - \sqrt{(\Pi_{1}(\Pi_{1} -1)}$ is negative, and hence, infeasible. Thus, asymptotic optimal PA  minimizing $\hat s_{o1}$ is given as \eqref{optimal_a_asy_sop1}.

\section{Proof of Lemma 5 }
Here also by setting, $\frac{\mathrm{d}\hat s_{o2}}{\mathrm{d}\alpha} = 0$ we obtain $\hat \alpha_{2} =  \Pi_{2} \pm \sqrt{(\Pi_{2}(\Pi_{2} -1)}$. We note $\hat \alpha_{2} = \Pi_{2} + \sqrt{(\Pi_{2}(\Pi_{2} -1)}$ is infeasible, because it forces $R_{s2}^{th}<0$ which is not possible. Hence, asymptotic optimal PA for U$2$ is given as \eqref{optimal_a_asy_sop2}.
\end{appendices}

\bibliographystyle{IEEEtran}
\bibliography{ref}

\end{document}